\documentclass[aps,prl,twocolumn,showpacs,floatfix]{revtex4}
\usepackage{graphicx}
\begin{document}
\newcommand{\App}{A^{\prime\prime}}
\title{Molecular collisions in ultracold atomic gases}
\author{Jeremy M. Hutson}
\affiliation{Department of Chemistry, University of Durham, South
Road, Durham, DH1~3LE, England}
\author{Pavel Sold\'{a}n}
\affiliation{Doppler Institute,Department of Physics, Faculty of
Nuclear Sciences and Physical Engineering, Czech Technical
University, B\v{r}ehov\'{a} 7, 115 19 Praha 1, Czech Republic}

\date{10 October 2006.
To be published in International Reviews in Physical Chemistry,
vol. 26, issue 1 (January 2007)}

\begin{abstract}
It has recently become possible to form molecules in ultracold
gases of trapped alkali metal atoms. Once formed, the molecules
may undergo elastic, inelastic and reactive collisions. Inelastic
and reactive collisions are particularly important because they
release kinetic energy and eject atoms and molecules from the
trap. The theory needed to handle such collisions is presented and
recent quantum dynamics calculations on ultracold atom-diatom
collisions of spin-polarised Li + Li$_2$, Na + Na$_2$ and K +
K$_2$ are described. All these systems have potential energy
surfaces on which barrierless atom exchange reactions can occur,
and both inelastic and reactive rates are very fast (typically
$k_{\rm inel} > 10^{-10}$ cm$^3$ s$^{-1}$ in the Wigner regime).
\end{abstract}


\maketitle

\tableofcontents

\section{Introduction}

There have been enormous recent advances in our ability to produce
and trap samples of cold molecules (below 1 K) and ultracold
molecules (below 1 mK). Molecules such as NH$_3$, OH and NH have
been cooled from room temperature to the milliKelvin regime by a
variety of methods including buffer-gas cooling
\cite{Weinstein:CaH:1998, Egorov:2004} and Stark deceleration
\cite{Bethlem:IRPC:2003, Bethlem:2006}. Molecules have also been
produced in ultracold atomic gases by photoassociation
\cite{Hutson:IRPC:2006, Jones:RMP:2006} and magnetoassociation
\cite{Hutson:IRPC:2006, Koehler:RMP:2006} of pairs of atoms.
Long-lived molecular Bose-Einstein condensates have been produced
for dimers of fermionic alkali metal atoms
\cite{Jochim:Li2BEC:2003, Zwierlein:2003, Greiner:2003}, and the
first signatures of ultracold triatomic \cite{Kraemer:2006} and
tetraatomic \cite{Chin:2005} molecules have been observed.

Cold and ultracold molecules have many possible applications.
High-resolution spectroscopy on cold molecules may allow the
measurement of fundamental physical properties such as the
electric dipole moment of the electron \cite{Hudson:2002}, the
energy differences between enantiomers \cite{Quack:2005,
Crassous:2005} and the time-dependence of the fine-structure
constant \cite{Hudson:2006}. In addition, since molecules have a
much richer energy level structure than atoms, they offer many new
possibilities for quantum control. Perhaps most importantly, {\it
dipolar} molecules interact with one another much more strongly
and at longer range than atoms. Dipolar molecules have been
proposed as qubits for quantum computers \cite{DeMille:2002} and
dipolar quantum gases are predicted to exhibit a range of novel
features \cite{Baranov:2002}.

In a recent article \cite{Hutson:IRPC:2006}, we reviewed the
current state of the art of molecule production in ultracold
atomic gases. Other authors have reviewed the cooling of molecules
from near room temperature \cite{Bethlem:IRPC:2003, Bethlem:2006}
and the theory of collisions of such directly cooled molecules
\cite{Krems:IRPC:2005, Bodo:IRPC:2006, Weck:IRPC:2006}. The
present article is complementary to these and focusses on recent
theoretical work on the collisions of alkali metal dimers formed
in ultracold gases.

\section{Experiments on ultracold molecule formation and collisions}

There are two main methods used to form molecules in ultracold
atomic gases. In {\em photoassociation} \cite{Hutson:IRPC:2006,
Jones:RMP:2006}, a pair of atoms undergoes a spectroscopic
transition from an unbound atomic state very close to threshold to
a bound molecular state. In {\em magnetoassociation}, also known
as {\em Feshbach resonance tuning} \cite{Hutson:IRPC:2006,
Koehler:RMP:2006}, the transition is accomplished by guiding the
atom pair adiabatically across an avoided crossing between an
unbound state and a molecular state.

Feshbach resonance tuning always produces molecules in very highly
excited states, usually the highest vibrational state that exists
in the diatomic potential well. Photoassociation also usually
produces molecules in very high vibrational states, because they
are the ones that have the best Franck-Condon overlap with the
free-atom states.

Ultracold molecules are initially formed in the presence of
ultracold atoms and can collide with them. For molecules in
vibrationally excited states, there is the possibility of
vibrationally inelastic collisions,
\begin{equation} \hbox{M}_2(v) + \hbox{M} \longrightarrow
\hbox{M}_2(v^\prime<v) + \hbox{M}, \end{equation} where $v$ is the
vibrational quantum number. Since the trap depth is usually much
less than 1 K, such collisions always release enough kinetic
energy to eject both collision partners from the trap. If the
molecular density is high, there is also the possibility of
inelastic molecule-molecule collisions,
\begin{equation} \hbox{M}_2(v) + \hbox{M}_2(v) \longrightarrow
\hbox{M}_2(v^\prime<v) + \hbox{M}_2 (v^{\prime\prime}\le v).
\end{equation}
Molecules are not {\it destroyed} in inelastic collisions, but
they are lost from the trap and are no longer ultracold.

The initial experiments on molecule formation by Feshbach
resonance tuning worked with bosonic isotopes of alkali metals
\cite{Donley:2002, Herbig:2003, Xu:2003, Durr:mol87Rb:2004}. They
found that the molecules were lost from the trap on a timescale of
milliseconds. The loss was attributed to vibrationally inelastic
atom-molecule collisions with relaxation rates around $10^{-10}$
cm$^3$ s$^{-1}$. For the case of $^{85}$Rb$_2$ \cite{Donley:2002},
recent work has suggested that the loss may in fact be due to
spontaneous molecular dissociation by collisionless spin
relaxation \cite{Thompson:spont:2005, Kohler:2005}. However, for
the other systems \cite{Herbig:2003, Xu:2003, Durr:mol87Rb:2004}
the molecules are formed in truly bound states and cannot decay
without collisions. Mukaiyama {\em et al.}\ \cite{Mukaiyama:2004}
have recently measured the trap loss rate for $^{23}$Na$_2$
molecules formed by Feshbach resonance tuning and obtained an
atom-molecule rate coefficient $k_{\rm loss} = 5.1\times10^{-11}$
cm$^3$ s$^{-1}$ for molecules in the highest vibrational state.

{\em Fermion} dimers formed by Feshbach resonance tuning are a
very special case. In mid-2003, four groups independently reported
within a very short time that dimers of fermionic $^6$Li
\cite{Strecker:2003, Cubizolles:2003, Jochim:Li2pure:2003} and
$^{40}$K \cite{Regal:lifetime:2004} could be remarkably stable to
collisions. Cubizolles {\em et al.}\ \cite{Cubizolles:2003} and
Jochim {\em et al.}\ \cite{Jochim:Li2pure:2003} showed that the
lifetime was particularly large close to a Feshbach resonance,
where the scattering length is large and positive. By the end of
2003, three different groups \cite{Jochim:Li2BEC:2003,
Zwierlein:2003, Greiner:2003} had succeeded in creating long-lived
molecular Bose-Einstein condensates of fermion dimers.

Petrov {\em et al.}\ \cite{Petrov:2004, Petrov:suppress:2005}
analysed the stability of fermion dimers in terms of the
long-range form of the wavefunction. In the case where the
atom-atom scattering length $a$ is much larger than the range of
the atom-atom potential $r_e$, they showed that both atom-molecule
and molecule-molecule inelastic collision rates are suppressed by
Fermi statistics. However, their derivation applies only to
molecules that are in long-range states, with a wavefunction that
depends on the scattering length, $\chi(r) \sim \exp(-r/a)$. As
will be discussed in more detail below, Cvita\v s {\em et al.}\
\cite{Cvitas:bosefermi:2005} have shown computationally that there
is {\em no} systematic suppression of the atom-molecule inelastic
rate for fermion dimers in low-lying vibrational levels, even when
$a$ is large and positive.

Relaxation processes have also been studied for molecules formed
by photoassociation. Wynar {\em et al.}\ \cite{Wynar:2000} formed
$^{87}$Rb$_2$ molecules in the second-to-last vibrational level of
the ground excited state by stimulated Raman adiabatic passage
(STIRAP). They estimated an upper bound of $k_{\rm
loss}=8\times10^{-11}$ cm$^3$ s$^{-1}$ due to inelastic
atom-molecule collisions. Staanum {\em et al.}\
\cite{Staanum:2006} investigated inelastic collisions of
rovibrationally excited Cs$_2$ ($^3\Sigma_u^+$) in collisions with
Cs atoms in two different ranges of the vibrational quantum number
$v$ by monitoring trap loss of Cs$_2$. They obtained atom-molecule
rate coefficients close to $1.0\times10^{-10}$ cm$^3$ s$^{-1}$ for
both $v=4$ to 6 and $v=32$ to 47. Zahzam {\em et al.}\
\cite{Zahzam:2006} carried out similar work for different
rovibrational states of $^3\Sigma_u^+$, and also considered
molecules in the $^1\Sigma_g^+$ state and molecule-molecule
collisions. They obtained rate coefficients of $2.6\times10^{-11}$
cm$^3$ s$^{-1}$ and $1.0\times10^{-11}$ cm$^3$ s$^{-1}$ in the
atom-atom and atom-molecule cases respectively, both with quite
large error bounds.

Because of the collisional losses for excited vibrational levels,
intense efforts are under way to produce ultracold alkali metal
dimers in low-lying levels (and ultimately in the ground vibronic
state). The process of transferring molecules from atomic or
near-dissociation molecular states, with probability density at
long range, to low-lying states, with probability density near the
diatomic equilibrium distance $r_e$, is sometimes called
$r$-transfer. It can in principle be achieved either by using many
photons to accomplish the transfer in several stages
\cite{Jaksch:2002} or by using tailored ultrafast laser pulses
\cite{Koch:2004, Koch:2006, Salzmann:2006}. In favourable cases,
and particularly for heteronuclear alkali metal dimers
\cite{Wang:1998, Stwalley:2004}, it may be possible to produce
molecules in their vibrational ground state by two-photon
processes via excited states with mixed singlet and triplet
character. Sage {\em et al.}\ \cite{Sage:2005} have recently
succeeded in creating ultracold RbCs molecules
($T\approx100\,\mu$K) in their vibronic ground state using a
4-photon process. The production rate in the current experiments
is only 500 molecules/s, but work is under way to increase it.

Inspired by the recent progress in experimental techniques for
producing and studying ultracold molecules, we have collaborated
with Jean-Michel Launay and coworkers at the University of Rennes
to produce a series of theoretical studies \cite{Soldan:2002,
Soldan:2003, Quemener:2004, Cvitas:bosefermi:2005,
Cvitas:li3:2006, Cvitas:hetero:2005, Quemener:2005,
Cvitas:longrange:2006} on interactions and collisions between
spin-polarized alkali-metal atoms and molecules. In particular,
ultra-low-energy collisions were studied for Na + Na$_{2}$
\cite{Soldan:2002, Quemener:2004}, Li + Li$_{2}$ (isotopically
homonuclear \cite{Cvitas:bosefermi:2005, Cvitas:li3:2006} and
heteronuclear \cite{Cvitas:hetero:2005}), and K + K$_{2}$
\cite{Quemener:2005}. Collisions between alkali-metal dimers and
atoms present several new theoretical challenges that are not
present for collisions of stabler molecules. The remainder of this
review will focus on describing this work and drawing general
conclusions from it.

\section{General aspects of alkali metal dimer collisions}

The atoms in an ultracold quantum gas of S-state atoms are in
states labelled by the electron spin $s$, the nuclear spin $i$,
and the total angular momentum $f$. Since there is usually a
magnetic field present, the states are also labelled by $m_f$, the
projection of $f$ along the field. Of particular interest here are
{\it spin-stretched} states, where $f=f_{\rm max}=i+s$ and
$|m_f|=f$. From a theoretical point of view, collisions of atoms
and molecules in spin-stretched states are simpler than others,
because they take place entirely on high-spin potential energy
surfaces. For alkali metal atoms with $s=\frac{1}{2}$, these are
triplet curves ($S=1$) for atom-atom collisions and quartet
surfaces ($S=\frac{3}{2}$) for atom-molecule collisions and 3-body
recombination. For atoms in non-spin-stretched states, singlet
dimer curves and doublet trimer surfaces are also required. Our
work so far has focussed on spin-stretched states.

The lowest quartet state of the alkali metal trimers, which
correlates with ground-state atoms ($^2$S) and molecules in their
lowest triplet state ($^3\Sigma_u^+$), is designated $(1^{4}A')$.
At first sight it might be expected that three parallel electrons
in s orbitals would not form significant chemical bonds. If this
was the case, then {\it pairwise additivity} would be a good
approximation and the potential energy surface would be given by
\begin{equation}
\label{Eqad} V(r_{12},r_{23},r_{13}) \approx \sum_{i<j}^3 V_{\rm
dimer}(r_{ij}).
\end{equation}
The potential energy curve for the lowest triplet state is
reasonably well known for many of the alkali metal dimers
\cite{Halls:2001, Minaev:2005, Ivanov:2003, Ahmed:2005}.
Pairwise-additive model potentials have been extensively used in
theoretical studies of 3-body recombination \cite{Esry:1999,
Suno:2002, Suno:fermions:2003, Suno:njp:2003}.

As will be seen below, pairwise additivity is actually quite a
poor approximation for the alkali metals. Nevertheless, it gives
some insights into the energetics of atom-molecule collisions. If
the dimer well depth is $\varepsilon$ at an atom-atom distance
$r_e$, then for a pairwise-additive surface the energies of
various important arrangements of 3 atoms are \begin{itemize}
\item atom M well-separated from diatom M$_2$ with bond length
$r_e$, $$V({\bf r})=V(r_e,\infty,\infty)=-\varepsilon$$ \item
linear trimer M$_3$ with bond length $r_e$, $$V({\bf
r})=V(r_e,r_e,2r_e)\approx -2\varepsilon$$ \item triangular trimer
M$_3$ with bond length $r_e$, $$V({\bf
r})=V(r_e,r_e,r_e)=-3\varepsilon.$$
\end{itemize}
The topology of the lowest quartet potential energy surface of an
alkali-metal trimer is thus quite simple; the global minimum is at
an equilateral triangular configuration (point group $D_{3h}$) and
there is a saddle point at a symmetric linear configuration
($D_{\infty h}$). The inclusion of nonadditivity deepens the
linear and (especially) triangular trimer wells, but does not
change the overall conclusions.

The potential energy surfaces are such that \textit{barrierless
atom-exchange reactions} can occur. {\it All} configurations of
the trimer M$_3$ are lower in energy than the separated atom +
diatom collision partners (and products). Once the collision
complex M$_3$ has been formed, any one of the three atoms can
depart to form products. Even when the products are
indistinguishable from the reactants, all three of these
``arrangement channels" must be taken into account in the
collision dynamics. This has two major consequences. First, a
reactive scattering approach (rather than an inelastic scattering
approach) must be used in the dynamics calculations. Secondly, the
scattering calculations must be done on a fully three-dimensional
potential energy surface. This is not the case for non-reactive
scattering, where the products and reactants are confined to just
one arrangement, and two-dimensional calculations (with the
diatomic bond length kept fixed at the dimer equilibrium geometry)
or quasi-three-dimensional calculations (where the diatomic bond
length is varied only slightly around the dimer equilibrium
geometry) are usually used.

The major focus of interest is in collisions that release kinetic
energy and thus lead to trap loss. These are typically vibrational
relaxation processes of the type
\begin{equation} \hbox{M}_2(v) + \hbox{M} \longrightarrow
\hbox{M}_2(v^\prime<v) + \hbox{M}. \end{equation} However, if the
three atoms are identical it is not possible to distinguish
between {\it inelastic} processes and {\it reactive} processes. We
therefore use the general term {\it quenching} to describe
collisions that produce a change in the vibrational (or
rotational) quantum number and release kinetic energy.

\section{Potential energy surfaces}

\subsection{Nonadditive forces}

Accurate quantum scattering calculations require accurate
potential energy surfaces. Higgins {\em et al.}\
\cite{Higgins:2000} showed in 2000 that the quartet state of
Na$_3$ exhibits strong non-additive forces that increase the well
depth of the equilateral trimer by 59\% and decrease the Na-Na
bond length from 5.2 \AA\ in the triplet dimer to 4.4 \AA\ in the
quartet trimer. We therefore carried out a systematic study
\cite{Soldan:2003} to investigate such effects for the whole
series of homonuclear alkali-metal trimers. \textit{Ab initio}
electronic structure calculations were performed using a
single-reference restricted open-shell variant \cite{Knowles:1993}
of the coupled-cluster method \cite{Cizek:1966} with single,
double and non-iterative triple excitations [RCCSD(T)].
Medium/large-size basis sets were used for the alkali metal atoms
as described in ref.\ \onlinecite{Soldan:2003}, and the full
counterpoise correction of Boys and Bernardi \cite{Boys:1970} was
employed to compensate for basis-set superposition errors. All the
\textit{ab initio} calculations were performed using the MOLPRO
package \cite{MOLPRO}.

\begin{figure} [htbp]
\begin{center}
\includegraphics[width=60mm,angle=-90]{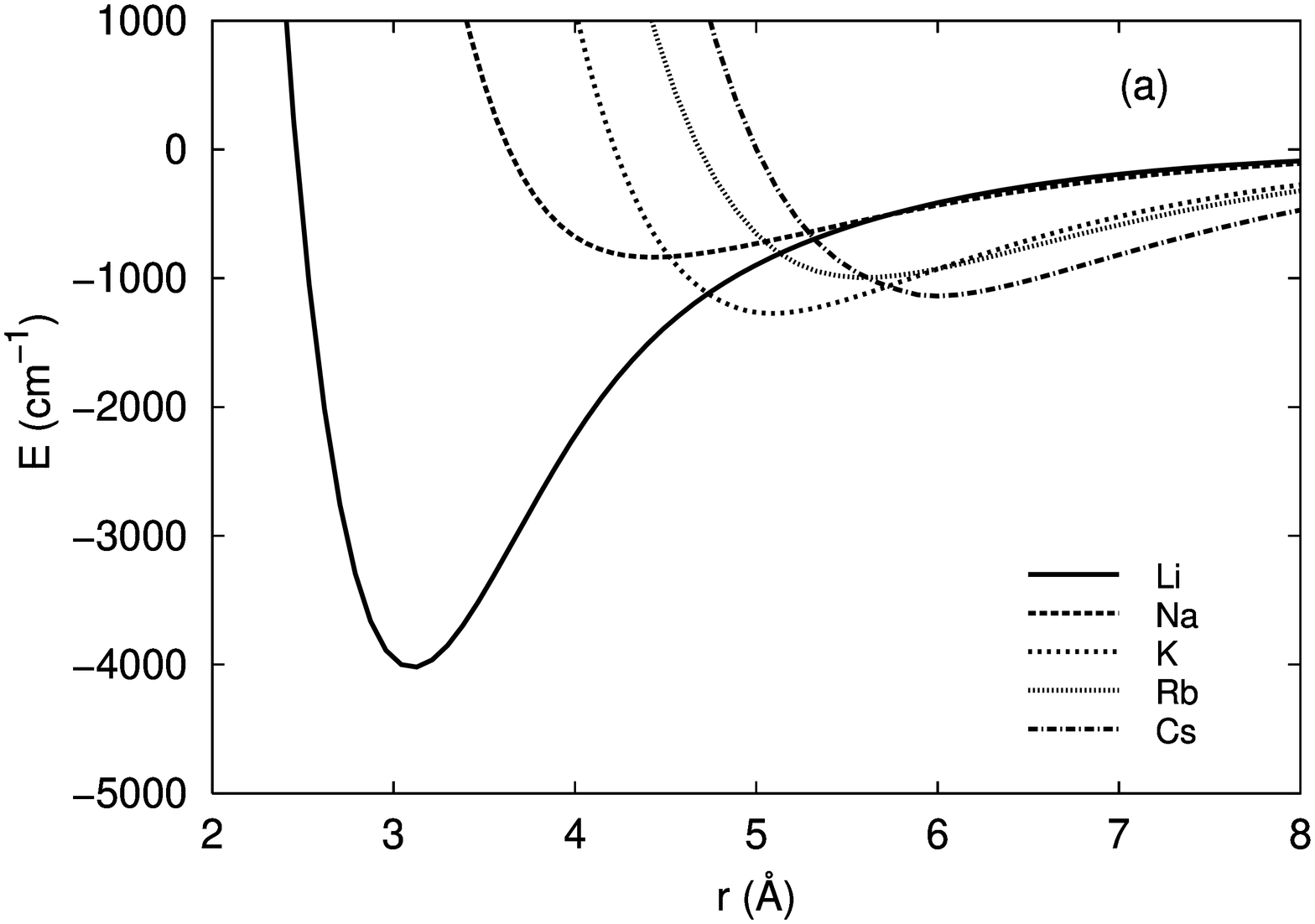}
\includegraphics[width=60mm,angle=-90]{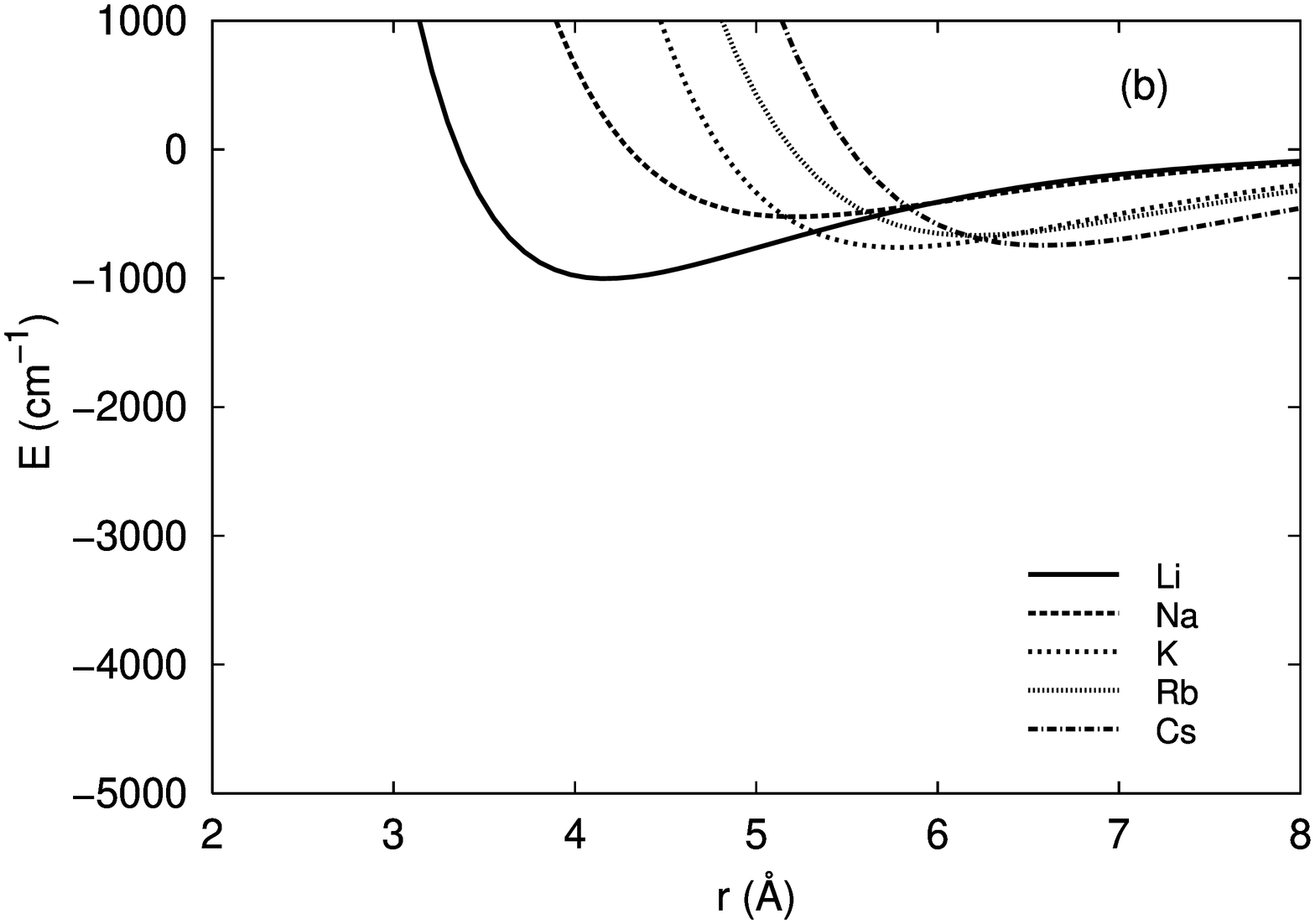}
\caption{RCCSD(T) interaction energies of spin-polarized alkali
trimers at $D_{3h}$ geometries (a) full potentials including
non-additive contributions; (b) additive potentials. Reproduced
from Sold\'an {\em et al.}\ \cite{Soldan:2003}.} \label{nonadd}
\end{center}
\end{figure}

The results are summarized in Table \ref{tnonadd}, which shows the
equilibrium bond lengths $r_{\rm e}$ and potential depths $V_{\rm
min}$ for alkali-metal dimers and equilateral ($D_{3h}$) trimers,
together with the corresponding quantities $r_{\rm sp}$ and
$V_{\rm sp}$ for the linear ($D_{\infty h}$) saddle points. The
three-body non-additive contributions $V_3$ are also given. Fig.\
\ref{nonadd} shows the additive and nonadditive potential energy
curves for $D_{3h}$ geometries.

\begin{table}[tb]
\caption{RCCSD(T) values of $r_{\rm e}$ (\AA), $r_{\rm sp}$ (\AA),
$V_{\rm min}=-D_{\rm e}$ (cm$^{-1}$), $V_{\rm sp}$ (cm$^{-1}$),
and $V_3$ (cm$^{-1}$) for spin-polarized alkali dimers and
trimers.} \label{tnonadd}
\begin{ruledtabular}
\begin{tabular}{lrrrrrrrr}
 & \multicolumn{2}{c}{Dimer} & \multicolumn{3}{c}{Trimer $D_{3h}$} &
\multicolumn{3}{c}{Trimer $D_{\infty h}$} \\
\cline{2-3} \cline{4-6} \cline{7-9}
 & $r_{\rm e}$ & $V_{\rm min}$ & $r_{\rm e}$ & $V_{\rm min}$ &  $V_{\rm 3}$ & $r_{\rm sp}$ & $V_{\rm sp}$ &  $V_{3}$ \\
\tableline
Li & 4.169 & -334.046 & 3.103 & -4022 & -5260 & 3.78 & -968 & -354 \\
Na & 5.214 & -174.025 & 4.428 &  -837 &  -663 & 5.10 & -381 &  -27 \\
K  & 5.786 & -252.567 & 5.084 & -1274 &  -831 & 5.67 & -569 &  -52 \\
Rb & 6.208 & -221.399 & 5.596 &  -995 &  -513 & 6.13 & -483 &  -15 \\
Cs & 6.581 & -246.786 & 5.992 & -1139 &  -562 & 6.52 & -536 &  -32 \\
\end{tabular}
\end{ruledtabular}
\end{table}

All the trimers show quite strong nonadditive effects. The quartet
trimers all have equilibrium interatomic distances (at $D_{3h}$
geometries) that are substantially shorter than those of the
triplet dimers, by an amount that decreases down the series from
1.07 \AA\ in Li$_3$ to 0.59 \AA\ in Cs$_3$. The trimer potentials
are all correspondingly deeper than pairwise sums of dimer
potentials, by a factor of 1.3 to 1.5 for the heavier alkali
metals (Na to Cs) but a factor of more than 4 for Li.

The size of the nonadditivity is at first sight quite surprising
in chemical terms. It contrasts with the situation for the rare
gas trimers, where the non-additive contributions are only 0.5\%
to 2.5\% \cite{Roeggen:1995, Lotrich:1997} and produce a {\it
weakening} of the binding at equilateral geometries rather than a
{\it strengthening} as in Table \ref{tnonadd}.

The interaction potentials can be decomposed into self-consistent
field (SCF) and correlation contributions. For the triplet alkali
dimers, as for the rare gas dimers, the SCF potentials are
repulsive and the main attractive forces arise from interatomic
correlation (dispersion). However, this similarity does {\it not}
extend to the trimers. For the rare gases, most of the
nonadditivity comes from the dispersion interaction. The leading
long-range term in this is the Axilrod-Teller-Muto (ATM)
triple-dipole term \cite{Axilrod:1943, Muto:1943}, which is
repulsive near equilateral configurations but attractive near
linear configurations. For alkali metal atoms, by contrast, there
is a large attractive contribution to the non-additive energy that
exists even at the SCF level. This occurs because the alkali
metals have vacant $n$p orbitals that lie relatively close to the
$n$s orbitals.

The question then arises why the p orbitals contribute so strongly
for the alkali metal trimers but not the dimers. Sold\'an {\em et
al.}\ \cite{Soldan:2003} carried out a natural orbital analysis
for equilateral triangle geometries, and considered the
contribution from {\it radial} p orbitals (pointing towards the
centre of the triangle) and from {\it tangential} p orbitals
(pointing around the ring). The radial p orbitals can form bonding
and antibonding molecular orbitals (MOs) of the same symmetry as
those formed from the $n$s orbitals ($a_1^\prime$ and $e^\prime$),
while the tangential p orbitals can form $a_2^\prime$ and
$e^\prime$ MOs. The sets of MOs of the same symmetry interact,
lowering the energy of the occupied MOs and contributing to
bonding. Sold\'an {\em et al.}\ showed that the dominant
contribution to the trimer bonding is from the tangential p
orbitals. In chemical terms, this is essentially sp hybridization.
This mechanism does not occur for the alkali metal dimers, because
in that case the ``tangential" p orbitals form molecular orbitals
of $\pi$ symmetry that cannot mix with $\sigma$ orbitals.

\subsection{Global potential energy surfaces}

It is relatively straightforward to generate potential energy
surfaces for systems such as the quartet alkali metal trimers by
carrying out {\it ab initio} electronic structure calculations at
grids of points that sample the configuration space. We have
carried out such calculations for Li$_3$
\cite{Cvitas:bosefermi:2005, Cvitas:li3:2006} and K$_3$
\cite{Quemener:2005}, using RCCSD(T) calculations as described
above for the linear and equilateral geometries. Colavecchia {\em
et al.}\ \cite{Colavecchia:2003} have also carried out
calculations for Li$_3$, using a more complete treatment of the
valence electron correlation (full configuration interaction) but
without correlating the core electrons.

A much more difficult problem is to generate a global potential
energy surface from a set of points. For quantum dynamics
calculations, it is very important to represent the
potential-energy function smoothly and without oscillations
between \textit{ab initio} points. If the resulting potential is
to be capable of representing all the properties of experimental
interest (including atom-atom scattering lengths, dimer and trimer
bound states, atom-diatom collisions and 3-body recombination),
then it is very important that the potential should dissociate
properly into all possible sets of products (atom + diatom and 3
separated atoms) with the correct long-range behaviour.

There are several coordinate systems that can be used for
triatomic systems, including hyperspherical coordinates, Jacobi
coordinates, and bond-length coordinates. These are by no means
equivalent for interpolation purposes. In particular, grids of
points in hyperspherical coordinates tend to include points in
which 2 atoms lie very close together, which hinders interpolation
because polynomials with very high localised maxima tend to have
oscillations in other regions. Jacobi coordinates suffer from the
same problem, and also do not allow the full 3-body exchange
symmetry to be introduced in a natural way. Fortunately, there is
no need to represent the potential energy surface in the same
coordinate system as is used in the dynamical calculations. We
therefore chose to carry out electronic structure calculations on
a grid of points in bond-length coordinates
$(r_{12},r_{23},r_{13})$. We use the shorthand $({\bf r})$ for
this to simplify notation.

In order to represent the atom-diatom dissociation limits
correctly, it is essential to use a long-range representation in
which the triatomic potential is decomposed into a sum of additive
and non-additive contributions,
\begin{equation}
\label{Eqnonad} V({\bf r}) = \sum_{i<j}^3 V_{\rm dimer}(r_{ij}) +
V_{3}({\bf r}). \label{eqv3}
\end{equation}
Provided $V_3({\bf r})\rightarrow 0$ when any two of the atom-atom
distances become infinite, this guarantees that the correct
diatomic potential is recovered in the atom-diatom limit. However,
low-energy scattering is very sensitive to long-range forces, so
we also require that the atom-diatom dispersion coefficients and
their anisotropies are correctly reproduced. This requires careful
treatment of the long-range part of $V_3({\bf r})$.

An important point is that, for a pairwise-additive potential in
which the dimer potentials have the correct long-range form $-C_6
R^{-6}$, the atom-diatom $C_6$ coefficient is {\it isotropic}
(independent of Jacobi angle $\theta$). The anisotropy of the
atom-diatom $C_6$ coefficient comes entirely from non-additive
forces. Cvita\v s {\em et al.}\ \cite{Cvitas:longrange:2006} have
therefore investigated the relationships between 3-body dispersion
coefficients in the atom-diatom and atom-atom-atom representations
and derived formulae relating the atom-diatom $C_6$ and $C_8$
coefficients and their anisotropies to three-body coefficients
arising from triple-dipole \cite{Axilrod:1943, Muto:1943},
quadruple-dipole \cite{Bade:1957, Bade:1958} and higher-order
multipole \cite{Bell:1970, Doran:1971} terms.

A variety of representations can be used for interpolating dimer
potentials. We have used the reciprocal-power reproducing kernel
Hilbert space method (RP-RKHS) \cite{Ho:1996, Ho:2000}. With an
appropriate choice of parameters \cite{Soldan:2000}, this gives a
potential that has the correct $C_6$ and $C_8$ dispersion
coefficients.

\begin{figure} [tbp]
\begin{center}
\includegraphics[width=85mm,angle=-90,bb=51 55 573 570]{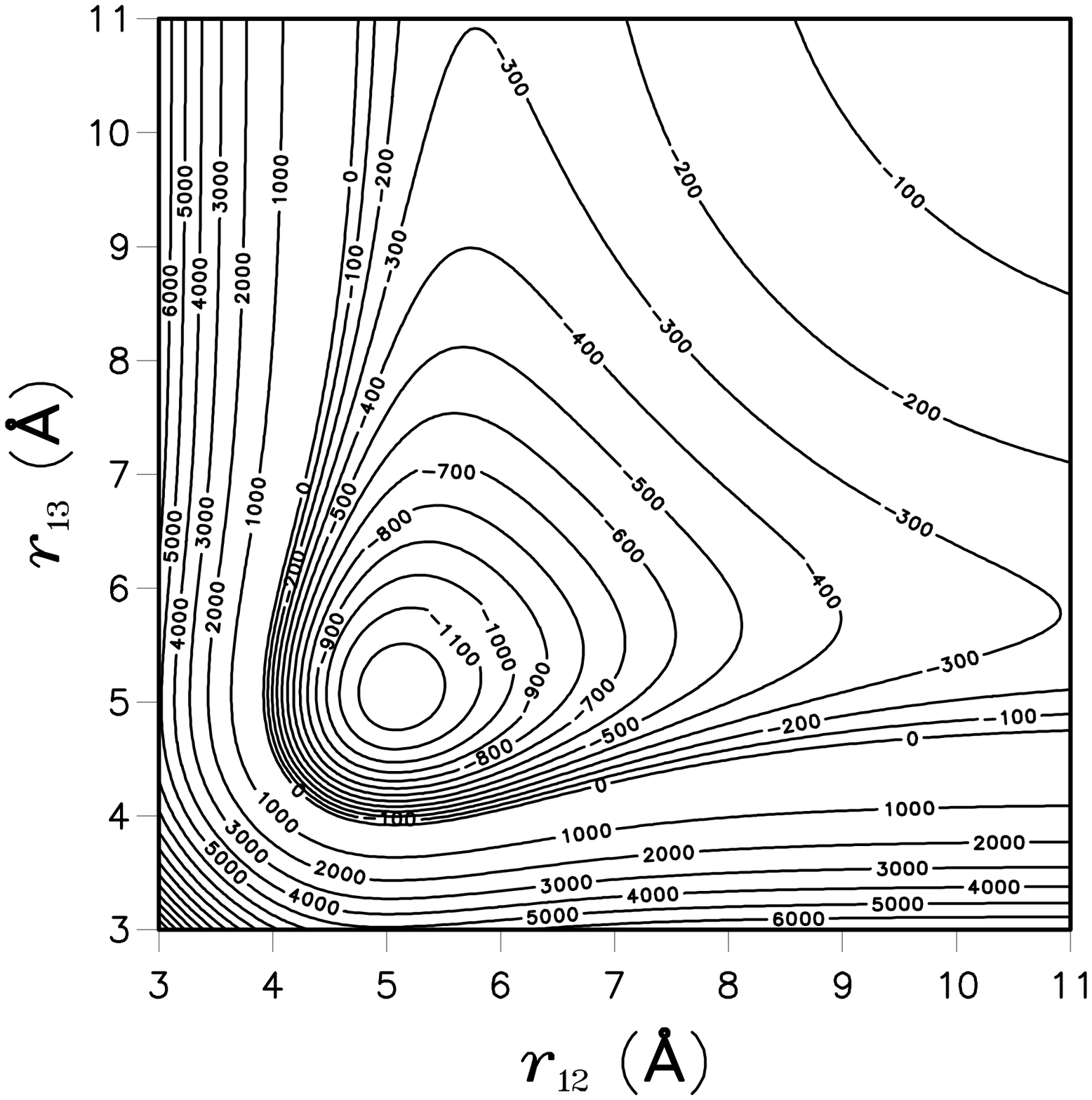}
\includegraphics[width=85mm,angle=-90,bb=51 55 573 570]{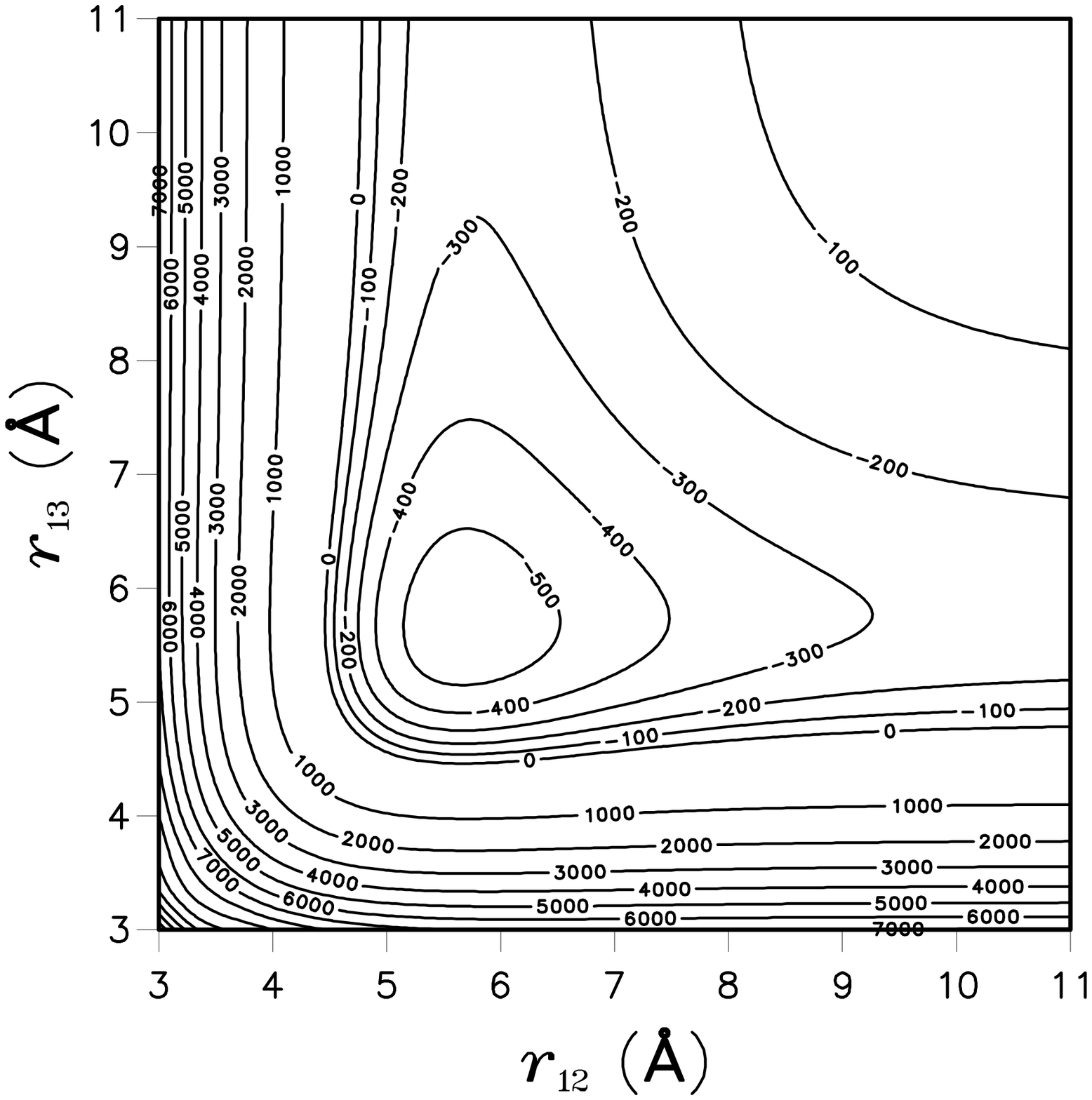}
\caption{Cuts through the K$_3$ quartet surface in valence
coordinates. Upper panel: cut for a bond angle of 60$^\circ$,
showing the global minimum at $-1269$ cm$^{-1}$ and 5.09 \AA.
Lower panel: cut at collinear geometries; the collinear minimum is
at $-565$ cm$^{-1}$ and 5.68 \AA. Contours are labelled in
cm$^{-1}$. Reproduced from Qu\'em\'ener {\em et al.}\
\cite{Quemener:2005}} \label{k3pot}
\end{center}
\end{figure}

At short range, different approaches were needed for K$_3$ and for
Li$_3$. K$_3$ is representative of the heavier alkali metals (Na
to Cs) in that the nonadditive potential is smaller than the
potential itself for most configurations. It is thus convenient to
use Eq.\ \ref{Eqnonad} directly with a global representation of
$V_3({\bf r})$. The details of the procedure are given in refs.\
\onlinecite{Quemener:2005} and \onlinecite{Cvitas:longrange:2006},
but in brief:
\begin{itemize} \item a quantity $V_3^\prime({\bf r})$ is defined by
subtracting out damped versions of the triple-dipole and
dipole-dipole-quadrupole terms from $V_3({\bf r})$,
\begin{equation} V_3'({\bf r}) = V_3({\bf r}) - \left[V^{\rm
DDD}_{\rm 3,damp}({\bf r}) + V^{\rm DDQ}_{\rm 3,damp}({\bf r})
\right];\end{equation} \item a function
\begin{equation} g({\bf r}) =
\frac{r_{12}^3r_{23}^3r_{13}^3}{(1+\cos^2\phi_{1})\,r_{23}^{6}+
(1+\cos^2\phi_{2})\,r_{13}^{6}+(1+\cos^2\phi_{3})\,r_{12}^6}
\end{equation} is defined to eliminate the quadruple-dipole contribution
to $V_3({\bf r})$; \item the function $V_3^{\prime\prime}({\bf r})
= g({\bf r}) \times V_3^\prime({\bf r})$ is interpolated using
3-dimension RP-RKHS interpolation. $V_3^{\prime\prime}({\bf r})$
is suitable for this (but $V_3({\bf r})$ and $V_3^\prime({\bf r})$
are not) because $V_3^{\prime\prime}({\bf r})$ takes a product
form at long range, $\hbox{constant} \times r_{12}^{-3}
r_{23}^{-3} r_{31}^{-3}$; \item $V_3({\bf r})$ is then rebuilt
from $V_3^{\prime\prime}({\bf r})$ at each interpolated point,
\begin{equation} V_3({\bf r}) = \frac{V_3''({\bf r})}{g({\bf r})}
+ \left[V^{\rm DDD}_{\rm 3,damp}({\bf r}) + V^{\rm DDQ}_{\rm
3,damp}({\bf r})\right].\end{equation}
\end{itemize}
The resulting potential energy surfaces for spin-polarized K$_3$
\cite{Quemener:2005} are shown in Fig.\ \ref{k3pot}. Note that the
depth at $D_{3h}$ geometries is rather more than twice that at
linear geometries, whereas pairwise additivity would give a factor
of 1.5.

Li$_3$ required a different procedure, because in this case the
potential minimum for the trimer occurs at a distance that is high
on the repulsive wall for the dimer. Because of this, Eq.\
\ref{Eqnonad} would represent the interaction potential in this
region as a difference between two very large quantities.
Nevertheless, at long range a decomposition according to Eq.\
(\ref{eqv3}) is essential. Under these circumstances, it is best
to fit the {\it ab initio} points directly to obtain a short-range
function $V_{\rm SR}({\bf r})$ without imposing the correct
long-range behaviour. A switching function $S({\bf r})$ is then
used to join this onto the correct long-range form,
\begin{equation}
V({\bf r}) = S({\bf r}) V_{\rm SR}({\bf r}) + [1-S({\bf r})]
V_{\rm LR}({\bf r}).
\end{equation}
The switching function is 1 at short range but switches smoothly
to zero at long range. The long-range form $V_{\rm LR}({\bf r})$
is designed to be valid when {\it any} of the atom-atom distances
is large, and the procedure used to build in the correct
three-atom and atom-diatom dispersion coefficients is described in
ref.\ \onlinecite{Cvitas:longrange:2006}.

\begin{figure} [tbp]
\begin{center}
\includegraphics[width=85mm,angle=-90]{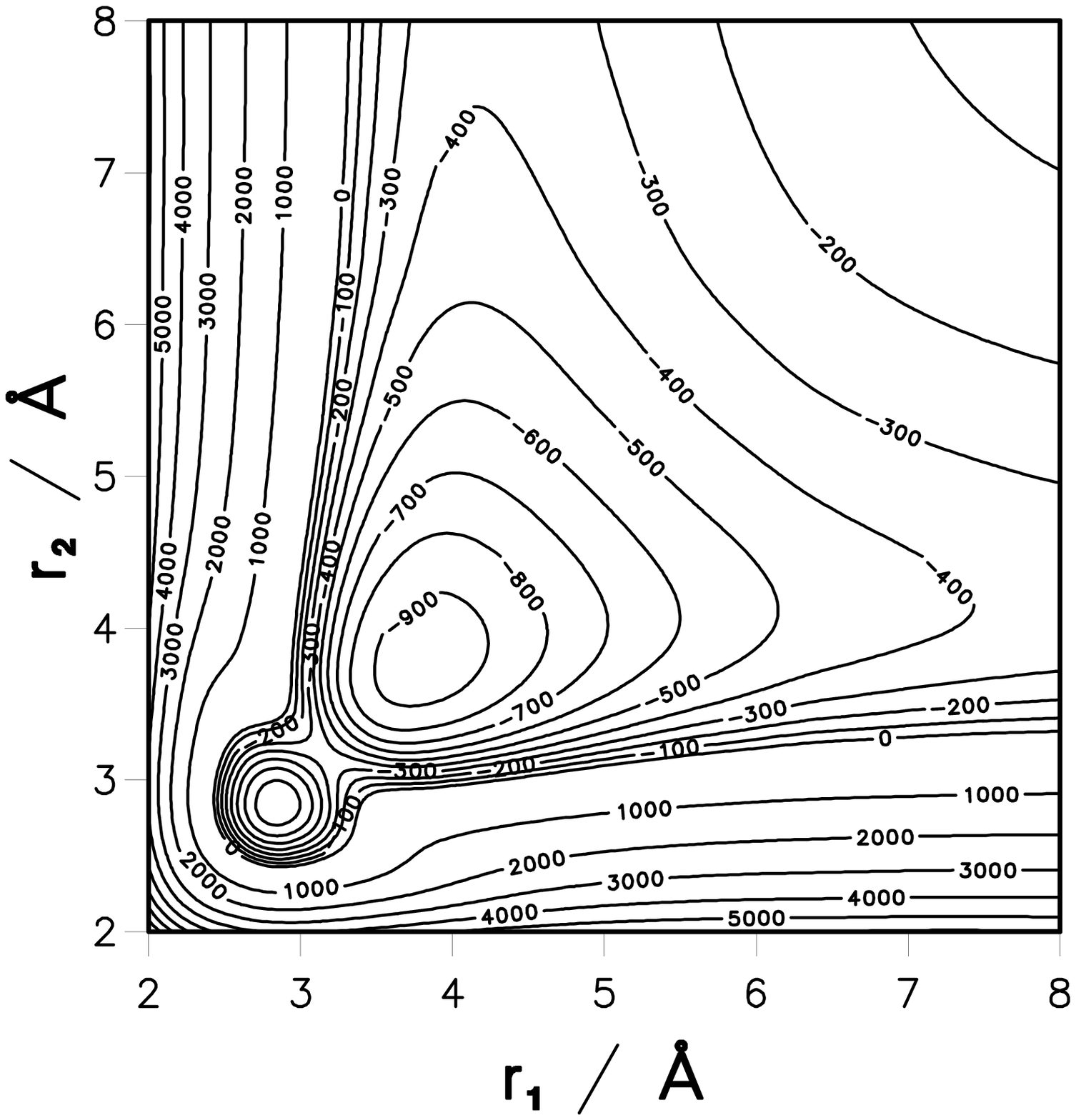}
\includegraphics[width=85mm,angle=-90]{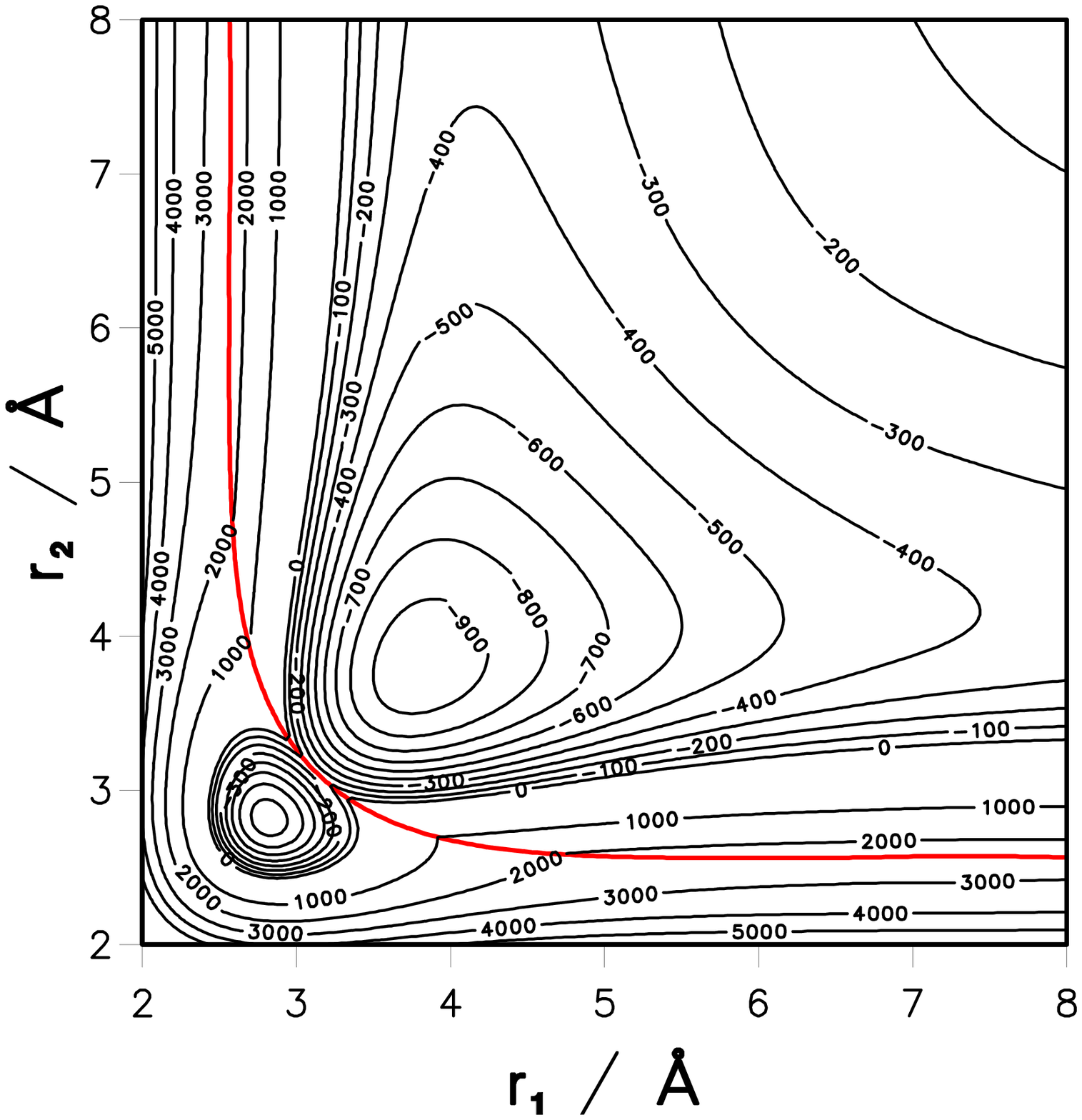}
\caption{Cuts through the Li$_3$ quartet surface in valence
coordinates. Upper panel: cut for a bond angle of 170$^\circ$,
showing the double-minimum structure due to avoided crossings at
near-linear geometries. Lower panel: cut at collinear geometries,
showing the seam of conical intersections (red line) between the
$^4\Sigma^+$ and $^4\Pi$ states. Reproduced from Cvita\v s {\em et
al.}\ \cite{Cvitas:bosefermi:2005}.} \label{Li3-pot}
\end{center}
\end{figure}

A further complication arises for Li$_3$ because in this case
there is a second potential energy surface involved. As described
above, the quartet state that correlates with 3 ground-state
($^2$S) atoms has $^4A'$ symmetry ($^4\Sigma^+$ at linear
geometries). The second state correlates with Li($^2$S) +
Li($^2$S) + Li($^2$P) and has $^4\Pi$ symmetry at linear
geometries. It can cross the $^4\Sigma^+$ state at linear
geometries because it has different symmetry. However, at
non-linear geometries the $^4\Pi$ state splits into $^4A'$ and
$^4A^{\prime\prime}$ components, and the two $^4A'$ states mix and
cannot cross. There are thus {\it conical intersections} at linear
geometries, and the avoided crossings produce double-minimum
structures at nonlinear geometries as shown in the upper panel of
Fig.\ \ref{Li3-pot}. The line of conical intersections produced a
{\it seam} in the potential surface as shown in the lower panel.

Conical intersections exist for both symmetrical and unsymmetrical
linear geometries in quartet Li$_3$, but for strongly
unsymmetrical geometries the conical intersections are high on the
repulsive wall and will not affect the dynamics. However, for
near-symmetric linear geometries the seam dips to an energy of
about $-100$ cm$^{-1}$ (relative to the energy of three free
atoms) at $r_{12}=r_{23}\approx 3.1$\,\AA. It is thus close to the
inner turning point for low-energy collisions between Li and
Li$_2$, and may have significant consequences for the chemical
dynamics. The conical intersection has subsequently been
characterized in more detail by Brue {\em et al.}\
\cite{Brue:2005}.

\section{Quantum dynamics calculations}

\subsection{Methodology}

As described above, alkali metal atom + diatom collisions require
a fully reactive scattering treatment, and at the energies of
interest for cold molecules it is essential to handle relative
translation as well as internal motions quantum-mechanically.
Quantum reactive scattering calculations \cite{Althorpe:2003} can
in general be performed using either {\it time-dependent} or {\it
time-independent} treatments. In recent work at higher energies,
time-dependent treatments based on wavepacket dynamics have been
becoming increasingly popular \cite{Althorpe:2004}. At ultralow
energies, however, the time evolution of a wavepacket is very
slow, and propagating it until it reaches the asymptotic region
requires an impractical number of time steps. Furthermore, it is
difficult to converge wavepacket calculations at very low
scattering energies, because the wave packet needs to be very
broad. We therefore chose to describe ultralow energy scattering
in a \textit{time-independent} formalism.

There are many variants of time-independent reactive scattering
theory. However, some of them are unsuitable for the alkali metal
trimers. As described above, the strong non-additive three-body
interactions for the alkali metal trimers make the atom-atom
distances at the trimer equilibrium geometry much shorter than
those for the dimers. The vibrational wavefunctions of a free
dimer are centred around its equilibrium bond length and are very
small at the distances that correspond to the trimer equilibrium.
Because of this, the free dimer wavefunctions do not form a good
basis set for expanding the scattering wavefunctions in the region
of the trimer equilibrium, where the actual atom exchange takes
place. This precludes the use of standard reactive scattering
packages such as the ABC program \cite{Skouteris:2000}, which uses
such basis functions and has been widely used in studies of
reactions such as F+H$_2$ \cite{Balakrishnan:FH2:2001,
Bodo:FD2:2002, Bodo:FH2:2004} and O($^3$P)+H$_2$
\cite{Weck:OH2:2005} at ultralow energies.

We thus chose to use a scattering formalism based on
hyperspherical coordinates $\rho,\theta,\phi$ \cite{Whitten:1968}:
$\rho$ is the hyperradius, which describes the size of the
triangle formed by the three atoms, while $\theta$ and $\phi$ are
hyperangles, which describe the shape of the triangle.
Hyperspherical approaches do not use free-diatom functions as a
basis set in the region of the trimer minimum. Instead, they
define an adiabatic basis set by solving a fixed-$\rho$
Schr\"odinger equation on a grid of values of the hyperradius
$\rho$. There are several different hyperspherical approaches
available. The approach developed for reactive scattering by Pack
and Parker \cite{Pack:1987, Parker:1987} solves the angular
problem using a finite-element approach in adiabatically adjusting
principal axis hyperspherical (APH) coordinates and then solves
the resulting radial coupled equations by propagation. The
approach developed by Esry and coworkers \cite{Esry:1996,
Suno:2002} and used extensively for 3-body recombination in cold
gases \cite{Suno:2002, Suno:fermions:2003, Suno:njp:2003} solves
the angular problem in slightly modified Smith-Whitten coordinates
\cite{Whitten:1968} using basis splines and then handles the
radial problem with a finite element approach. Both these methods
use an optimised nonuniform grid in the hyperangles. However, we
chose to use an alternative approach developed by Launay and
LeDourneuf \cite{Launay:1989}, which has been applied extensively
to chemical reactions such as N($^2$D)+H$_2$ \cite{Honvault:1999}
and O($^1$D)+H$_2$ \cite{Honvault:2001} at higher energies.

In the approach of Launay and LeDorneuf, the configuration space
is divided into inner and outer regions, and the boundary between
them is placed at a distance (hyperradius) such that couplings due
to the residual atom-diatom interaction can be neglected outside
the boundary. This distance is typically $\rho=45$ to 60 $a_0$ for
the alkali metal systems. In the inner region, the wavefunction
for nuclear motion is obtained by propagating a set of coupled
differential equations using a diabatic-by-sector algorithm. The
angular basis set is obtained by diagonalizing a fixed-hyperradius
reference Hamiltonian in a primitive basis set of
pseudohyperspherical harmonics. In the outer region, the
wavefunction is expanded in a basis set of diatom
vibration-rotation functions expressed in Jacobi coordinates
\cite{Arthurs:1960}. The wavefunctions in the outer region are
computed by inwards integration of regular and irregular solutions
of an uncoupled radial Schr{\"o}dinger equation which includes the
isotropic part of the atom-molecule interaction. Matching between
wavefunctions for the inner and outer regions yields the
scattering S-matrix. Elastic and inelastic cross sections are then
calculated using standard formulae \cite{Quemener:2005}.

The size of the basis set required for convergence depends
strongly on the masses involved and the depth of the potential
energy well. Before our calculations on alkali metal systems,
nearly all quantum scattering calculations had been on systems
containing only one or occasionally two non-hydrogen atoms.
Calculations on alkali metal atoms require much bigger angular
basis sets, though fortunately at low energies only a few partial
waves (values of $J$, the total angular momentum excluding spin)
are needed. Even for Li + Li$_2$ \cite{Cvitas:bosefermi:2005},
which is relatively light, the number of adiabatic angular
functions retained in the coupled equations ranged from 97 for
$J=0$ to 827 for $J=10$, while for K + K$_2$ \cite{Quemener:2005}
the range was from 250 for $J=0$ to 1411 for $J=5$.

One problem with hyperspherical methods is that the diatom
functions become more localised in hyperangular space as $\rho$
increases. Because of this, the number of hyperspherical harmonics
needed for convergence increases with $\rho$: for K + K$_2$ it
varied from 867 functions at small $\rho$ to 6625 functions at the
matching distance. It is the calculation to build the adiabatic
angular basis that dominates the computer requirements (180 hours
on an IBM Power4 P960 for K + K$_2$).

A major advantage of the hyperspherical harmonics is that boson or
fermion symmetry is very easy to impose. The complete nuclear
permutation group for a system with three identical nuclei is
$S_{3}$. To satisfy the Pauli principle, the total wavefunction
must have $A_1$ symmetry for bosonic nuclei or $A_2$ symmetry for
fermionic nuclei. The total wavefunction is in general a sum of
products of electronic, nuclear spin and nuclear motion parts. For
three atoms in their spin-stretched states, the nuclear spin
wavefunction is totally symmetric. For such states, collisions
take place entirely on the quartet surface, for which the
electronic wavefunction has $A_2$ symmetry. Boson or fermion
symmetry can thus be imposed by selecting pseudohyperspherical
harmonics to give the correct symmetry for the wavefunction for
nuclear motion. The adiabatic states in each sector are obtained
by a variational expansion on a basis of hyperspherical harmonics
with $A_1$ symmetry for bosonic atoms (with fermionic nuclei) and
$A_2$ symmetry for fermionic atoms (with bosonic nuclei).

All the calculations were carried out on the quartet trimer
surfaces, so are appropriate for collisions of spin-polarised
atoms and molecules. However, the basis functions used for the
quantum dynamics calculations did not explicitly include electron
spin. In a more complete treatment, the rotational quantum number
$n$ for the triplet dimer would couple to its spin $s=1$ to give a
resultant $j$. When spin is neglected, however, there is no
distinction between $n$ and $j$. The splittings between levels of
the same $n$ but different $j$ are in any case very small for the
alkali metal dimers.

\subsection{Homonuclear molecules}

Quantum dynamics calculations have been carried out for the
homonuclear collisions Li + Li$_2(v=0\hbox{ to }3)$
\cite{Cvitas:bosefermi:2005, Cvitas:li3:2006}, Na +
Na$_2(v=0\hbox{ to }3)$ \cite{Soldan:2002, Quemener:2004} and K +
K$_2(v=1)$ \cite{Quemener:2005}. For the Li and K systems,
calculations were carried out for both bosonic and fermionic
isotopes.

\begin{figure} [tbp]
\begin{center}
\includegraphics[width=67mm,angle=-90,bb=88 26 570 628]{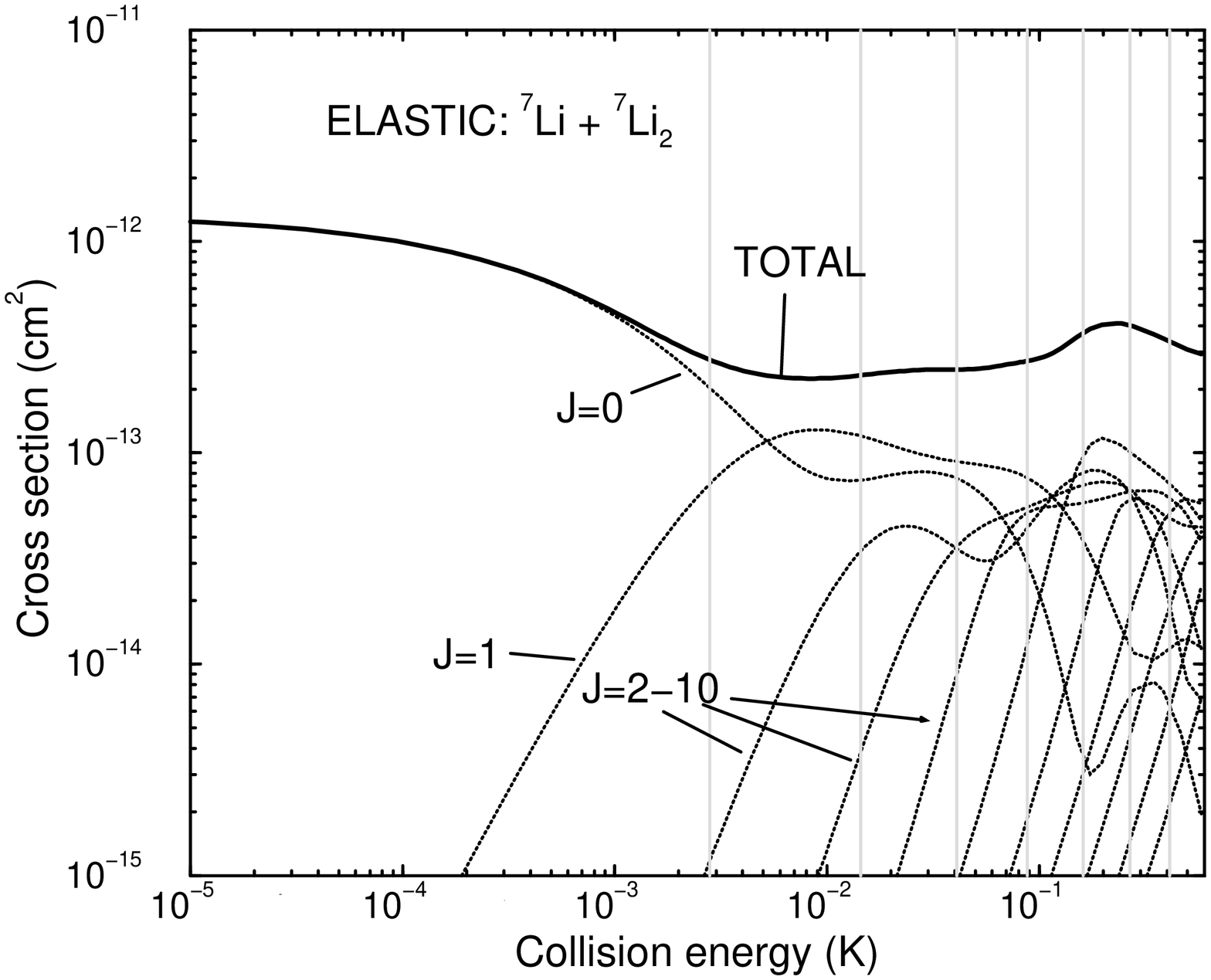}
\includegraphics[width=67mm,angle=-90,bb=88 26 570 628]{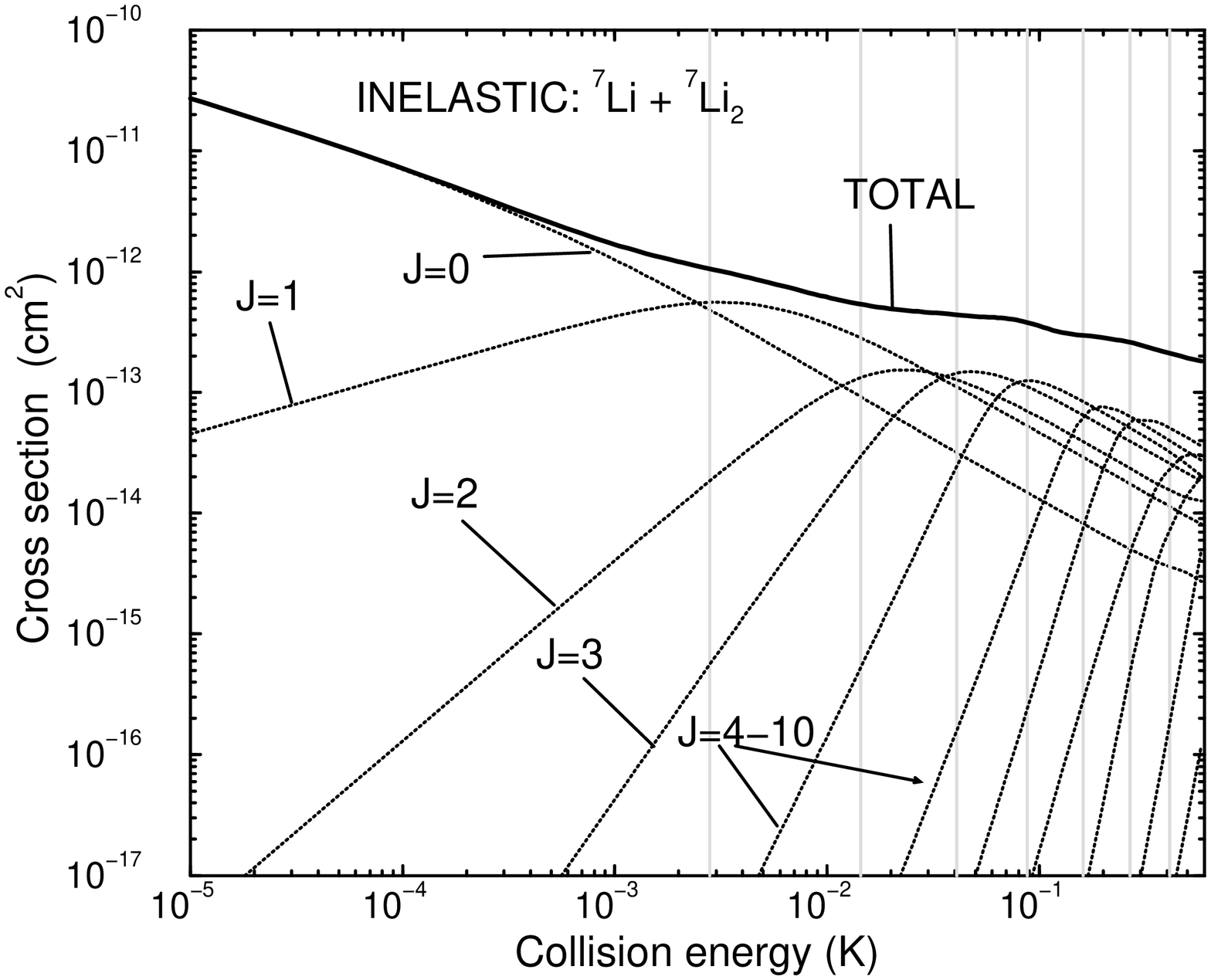}
\caption{Elastic cross sections (upper panel) and inelastic cross
sections (lower panel) for $^7$Li + $^7$Li$_2$($v_i=1$, $j_i=0$),
with contributions from individual partial waves (total angular
momentum $J$, excluding spin). The vertical lines indicate
centrifugal barrier heights for $l\ge1$. Reproduced from Cvita\v s
{\em et al.}\ \cite{Cvitas:bosefermi:2005}.} \label{liboson}
\end{center}
\end{figure}

\begin{figure} [tbp]
\begin{center}
\includegraphics[width=67mm,angle=-90,bb=88 26 570 628]{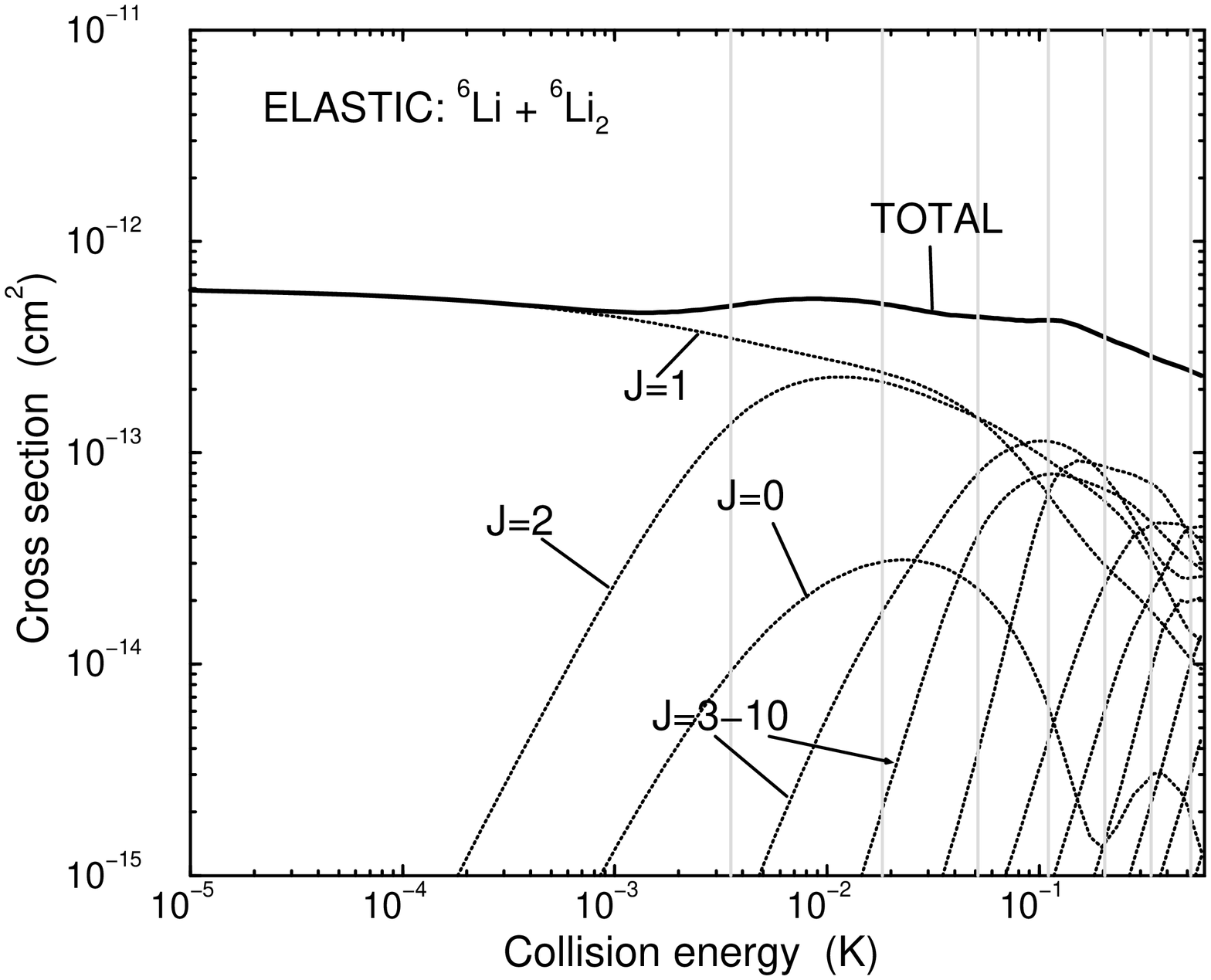}
\includegraphics[width=67mm,angle=-90,bb=88 26 570 628]{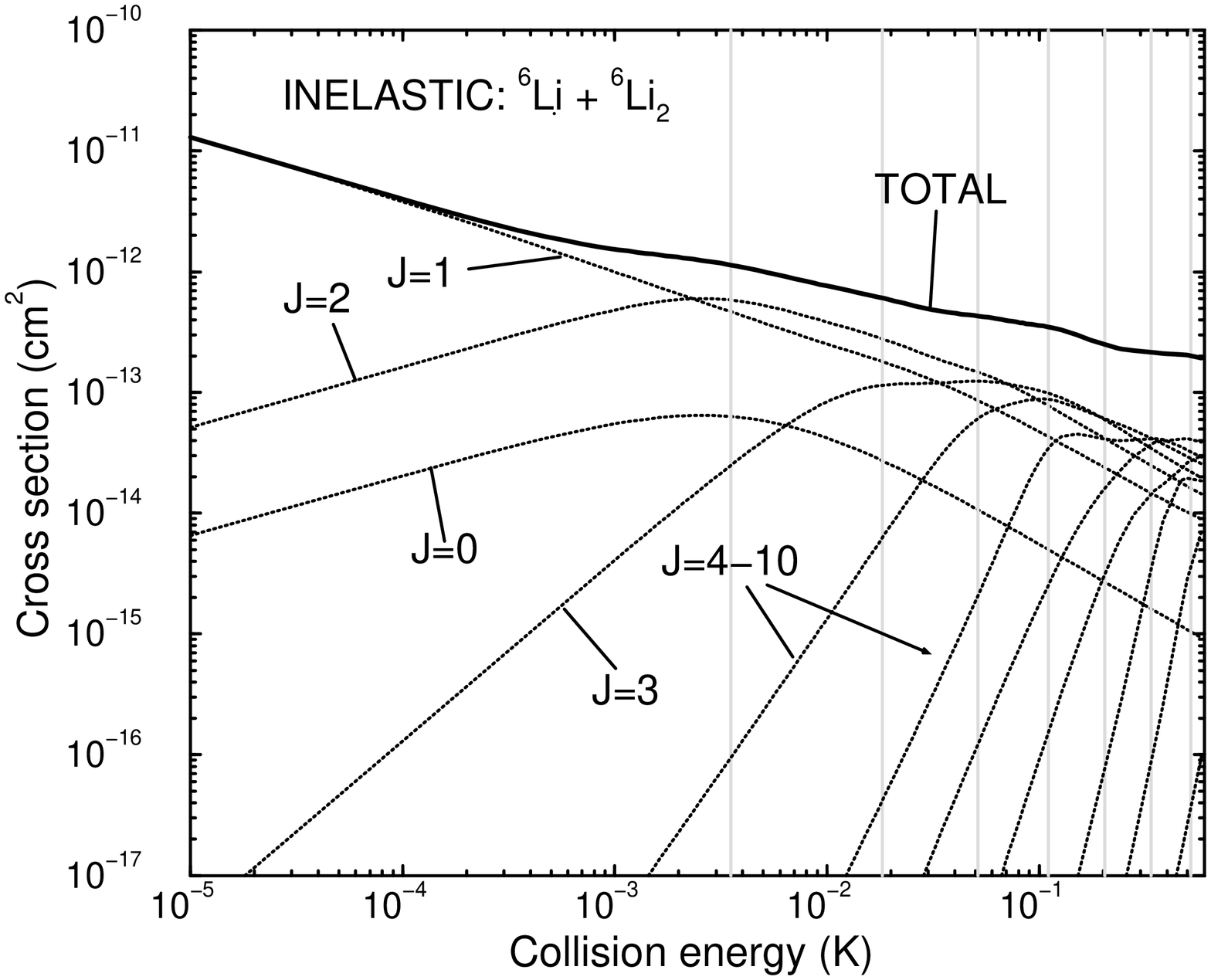}
\caption{Elastic cross sections (upper panel) and inelastic cross
sections (lower panel) for $^6$Li + $^6$Li$_2$($v_i=1$, $j_i=1$),
with contributions from individual partial waves (total angular
momentum $J$, excluding spin). The vertical lines indicate
centrifugal barrier heights for $l\ge1$. Reproduced from Cvita\v s
{\em et al.}\ \cite{Cvitas:bosefermi:2005}.} \label{lifermion}
\end{center}
\end{figure}

\begin{figure} [tbp]
\begin{center}
\includegraphics[width=67mm,angle=-90]{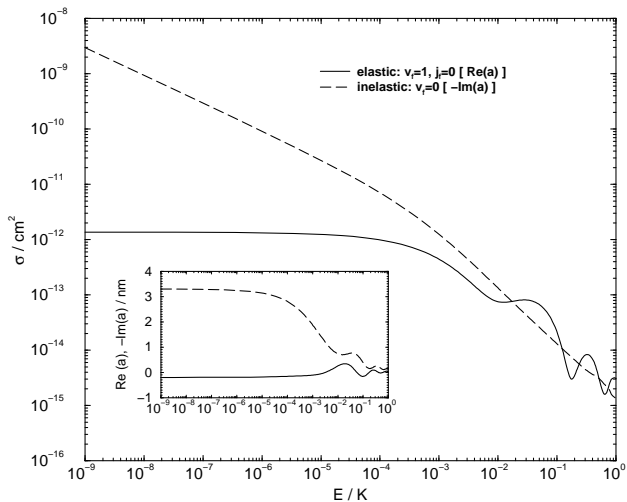}
\caption{Elastic and inelastic s-wave ($J=0$) cross sections for
$^7$Li + $^7$Li$_2$($v_i=1$, $j_i=0$). The inset shows the real
and imaginary parts of the complex scattering length.}
\label{li7-s-wave}
\end{center}
\end{figure}

The results for Li are typical. Figs.\ \ref{liboson} and
\ref{lifermion} show elastic and inelastic cross sections for
bosons ($^7$Li) and fermions ($^6$Li) respectively. The elastic
and inelastic s-wave cross sections for $^{7}$Li are compared
directly and extended to lower energy in Fig.\ \ref{li7-s-wave}.
It may be seen that at very low energies (below 100 $\mu$K for Li)
the elastic cross sections become independent of energy whereas
the inelastic cross sections are proportional to $E_{\rm
kin}^{-1/2}$. This is in accordance with the Wigner threshold
laws, which state that at very low energy the partial cross
sections (contributions from a single partial wave $l$) for
elastic and inelastic scattering vary as
\begin{equation}
\sigma_{\rm el}^l  \sim E_{\rm kin}^{2l}; \hskip 5mm \sigma_{\rm
inel}^l \sim E_{\rm kin}^{l-1/2}. \label{wcross}
\end{equation}
For a long-range potential proportional to $R^{-6}$, as for
neutral atom-diatom scattering, there is an $l$-independent term
that dominates the threshold law for high $l$ so that $\sigma_{\rm
el}^l \sim E_{\rm kin}^3$ for $l\ge2$ \cite{Sadeghpour:2000}.

It may be seen from Figs.\ \ref{liboson} and \ref{lifermion} that
below $E_{\rm kin}=100\,\mu$K the cross sections are completely
dominated by the $l=0$ term, which corresponds to total angular
momentum $J=0$ for bosons but $J=1$ for fermions (because $J=l+j$
and the lowest rotational level of a triplet fermion dimer is
$j=1$). The energy-dependence of the inelastic rate coefficient is
$k\sigma_{\rm inel}$ where $k=(2E_{\rm kin}/ \mu)^{1/2}$, so that
the inelastic rate is independent of energy in this region.

Above $E_{\rm kin}=100\,\mu$K, the $l=0$ contribution starts to
deviate from the Wigner limit and higher partial waves start to
contribute. The point at which this happens depends somewhat on
mass, and is closer to 10\,$\mu$K for K + K$_2$
\cite{Quemener:2005}.

As described in Section 2, it is known experimentally that fermion
dimers produced by Feshbach resonance tuning are very much stabler
than boson dimers when the scattering length is large. This
stability was crucial in the production of molecular Bose-Einstein
condensates of $^6$Li$_2$ \cite{Jochim:Li2BEC:2003,
Zwierlein:2003} and $^{40}$K$_2$ \cite{Greiner:2003}. Petrov {\em
et al.}\ \cite{Petrov:2004} explained the stability in terms of
the requirements of Fermi-Dirac statistics. However, their
derivation is valid only for long-range molecular states. A very
important question is whether the stability persists for low-lying
vibrational states of fermion dimers.

Cvita\v s {\em et al.}\ \cite{Cvitas:bosefermi:2005} carried out
quantum dynamics calculations for spin-polarized collisions of
ultracold homonuclear Li + Li$_2$ collisions for both the bosonic
($^7$Li) and fermionic ($^6$Li) cases. The results shown in Figs.\
\ref{liboson} and \ref{lifermion} correspond to vibrational
quenching rates for $v=1$ of $k_{\rm inel} = 5.6 \times 10^{-10}$
cm$^3$ s$^{-1}$ for bosons and $2.8 \times 10^{-10}$ cm$^3$
s$^{-1}$ for fermions. However, the apparent difference of a
factor of 2 is not significant: there were differences of up to a
factor of 8 between inelastic cross sections for different initial
values of $v$ and $j$. Cvita\v s {\em et al.}\ found no {\it
systematic} difference between inelastic rates in the boson and
fermion cases, even when the atom-atom scattering length was
adjusted to be large and positive. This has important consequences
for efforts to transfer the dimer population from Feshbach
resonance states to the vibrational ground state, $v=0$: it will
be necessary to carry out the process either in a single step or
quickly enough that the molecules do not spend enough time in
intermediate states to undergo inelastic collisions.

The effective potential for a partial wave with $l>0$ is governed
at long range by the centrifugal and dispersion terms,
\begin{eqnarray}
V^l(R)=\frac{\hbar^2 l(l+1)}{2 \mu R^2} - \frac{C_6}{R^6},
\end{eqnarray}
where $C_6$ is the atom-diatom dispersion coefficient. There is
thus a centrifugal barrier at a distance
\begin{eqnarray}
R_{\rm max}^l=\left[\frac{6 \mu C_6}{\hbar^2 l(l+1)}\right]^{1/4}
\end{eqnarray}
with height
\begin{eqnarray}
V_{\rm max}^l=\left[\frac{\hbar^2 l(l+1)}{\mu}\right]^{3/2}
(54C_6)^{-1/2}.
\end{eqnarray}

The resulting barrier heights are included in Figs.~\ref{liboson}
and \ref{lifermion}. The first vertical line corresponds to the
$l=1$ partial wave and so on up to the $l=7$ partial wave. It may
be seen that each partial cross section has a maximum at an energy
slightly higher than the corresponding $V_{\rm max}^l$. At
collision energies below the centrifugal barrier, the partial
cross sections for each $l$ follow Wigner laws given by Eq.\
(\ref{wcross}).  Above the centrifugal barrier, the inelastic
probabilities come close to their maximum value of 1 and the cross
sections vary as $E^{-1}$ because of the $k^{-2}$ factor in the
expression for the cross section.

\subsection{Capture model outside the ultracold regime}

\begin{figure} [htbp]
\begin{center}
\includegraphics[width=68mm,angle=-90,bb=89 28 574 630]{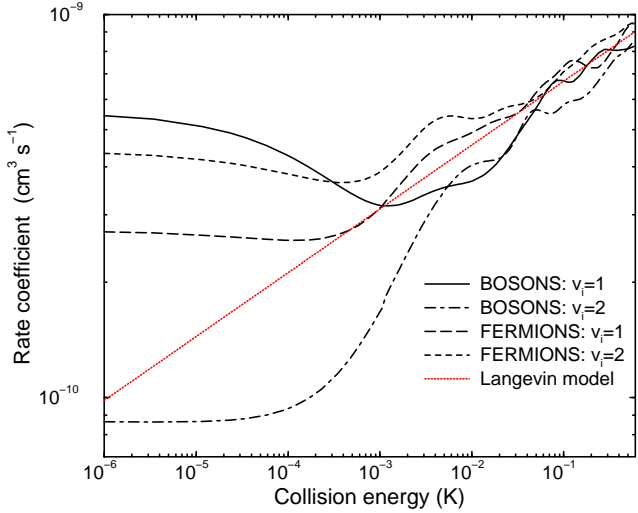}
\caption{Total inelastic rate coefficients for collisions of Li
with Li$_2$ ($v=1$ and 2, with $j=0$ for bosons and $j=1$ for
fermions). Reproduced from Cvita\v s {\em et al.}\
\cite{Cvitas:bosefermi:2005}.} \label{LiLangevin}
\end{center}
\end{figure}

At high collision energy, when several partial waves are involved,
the total inelastic rate coefficient can be compared with that
given by the classical Langevin capture model \cite{Levine:1987},
based on the idea that every collision that crosses the
centrifugal barrier produces inelasticity. This gives
\begin{eqnarray} \sigma_{\rm inel}^{\rm capture}(E) &=& 3\pi
\left(\frac{C_6}{4E}\right)^{1/3}; \cr k_{\rm inel}^{\rm
capture}(E) &=& 3\pi \left(\frac{C_6}{4E}\right)^{1/3}
\left(\frac{2E}{\mu}\right)^{1/2} = \frac{3\pi C_6^{1/3}
E^{1/6}}{2^{1/6}\mu^{1/2}}.
\end{eqnarray}
This rate coefficient is shown as a function of collision energy
for Li + Li$_2$ in Fig.\ \ref{LiLangevin} and compared with the
quenching rates for bosons and fermions initially in $v=1$ and 2
\cite{Cvitas:bosefermi:2005}. It may be seen that the full quantum
result approaches the Langevin value at collision energies above
about 10 mK. Similar behaviour is seen for K + K$_2$ at collision
energies above 0.1 mK \cite{Quemener:2005}.

\subsection{Product rotational distributions}

\begin{figure}[tbp]
\begin{center}
\includegraphics[width=65mm,angle=-90]{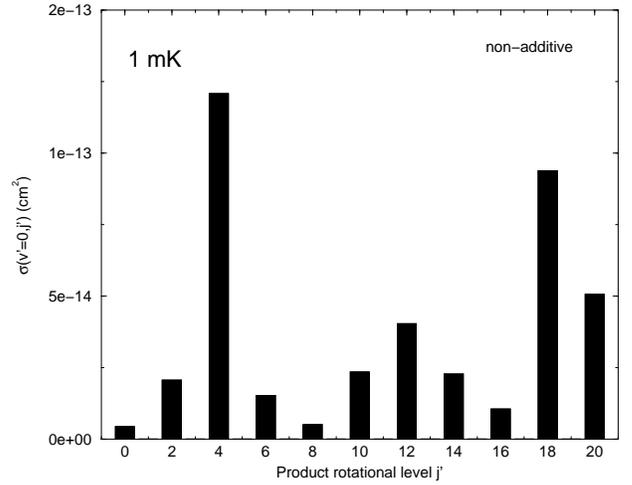}
\caption[Rotational distribution] {Rotational distributions for
$^{23}$Na + $^{23}$Na$_2\ (v=1)$ at collision energy $10^{-4}$ K.
The label $j'$ is the final rotational quantum number of
$^{23}$Na$_2\ (v'=0)$. Reproduced from Sold\'an {\em et al.}\
\cite{Soldan:2002}.} \label{na2rot}
\end{center}
\end{figure}

The vibrational spacings of the alkali metal dimers are much
larger than their rotational spacings, so that many rotational
levels are energetically accessible in collisions that cause
vibrational relaxation. For Na$_2$, for example, rotational levels
up to $j=20$ are energetically accessible at the energy of the
$v=1$ state (23.5 cm$^{-1}$). All accessible levels are populated
in the products, subject to symmetry restrictions (only even-$j$
levels for bosons and odd-$j$ levels for fermions). The product
rotational distributions for Na + Na$_2\ (v=1)$ at $10^{-4}$ K are
shown in Fig.\ \ref{na2rot}. There are three clear maxima in the
distribution, at $j=4$, 12 and 18. The oscillations probably arise
from a rotational rainbow effect \cite{Bowman:1979, Schinke:1979}.
The structure is similar to that observed in vibrational
predissociation of Van der Waals complexes
\cite{Hutson:ArH2:1983}. In a classical impulsive model, the
energy released from Na$_2$ vibration is partly retained in
relative translation and partly converted into Na$_2$ rotation.
The angular momentum imparted to the Na$_2$ molecule is zero if
the energy is released at a linear or T-shaped geometry, but large
around $\theta=45^\circ$. In this model, the oscillations arise
from interference between classical trajectories on either side of
the maximum.

\begin{figure}[tbp]
\begin{center}
\includegraphics[width=85mm,height=65mm]{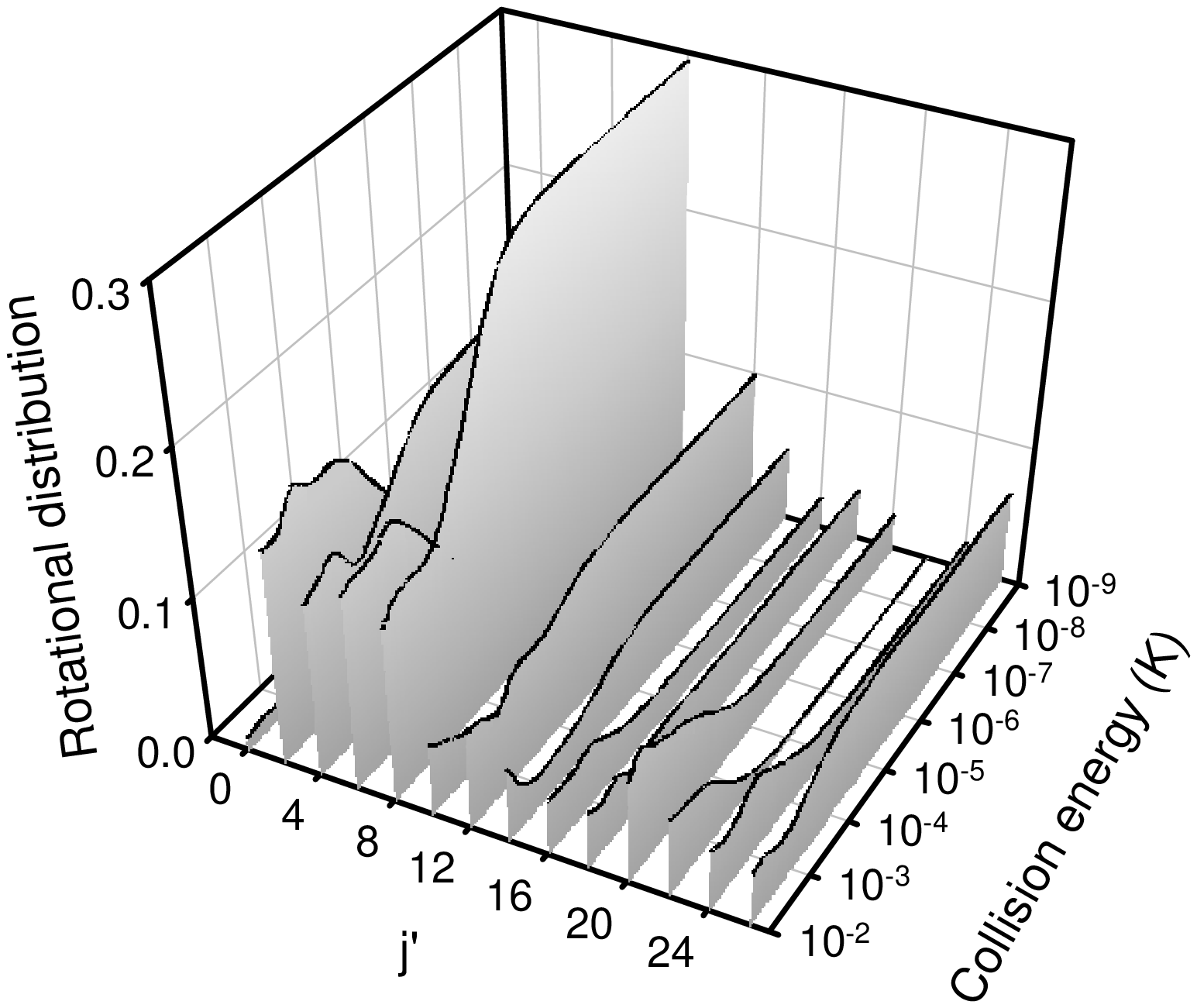}
\caption[Rotational distribution] {Rotational distributions for
$^{39}$K + $^{39}$K$_2\ (v=1)$ as a function of the collision
energy. The label $j'$ is the final rotational quantum number of
$^{39}$K$_2\ (v'=0)$. Reproduced from Qu\'em\'ener {\em et al.}\
\cite{Quemener:2005}.} \label{k2rot}
\end{center}
\end{figure}

The rotational distributions become constant at low energies but
show structure above $10^{-4}$ K. This is shown for K + K$_2\
(v=1)$ in Fig.\ \ref{k2rot}. In this case levels up to $j=24$ are
energetically accessible. Once again there is an oscillatory
structure in the product state distributions.

\subsection{Potential sensitivity}

\begin{figure} [tbp]
\begin{center}
\includegraphics[width=65mm,angle=-90]{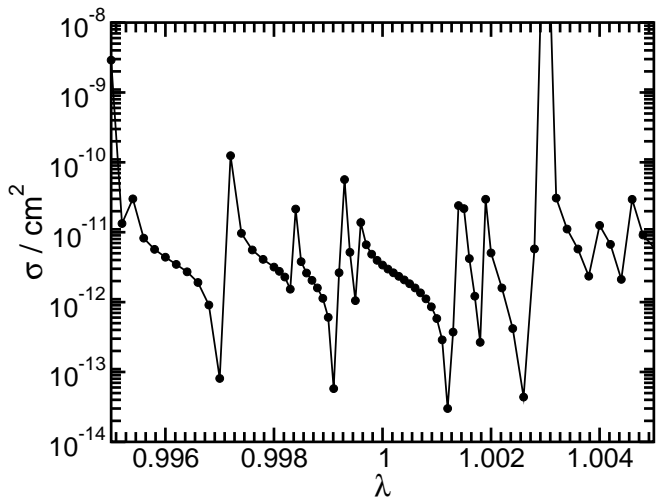}
\caption{Dependence of the elastic cross sections for
$^7\mbox{Li}+\mbox{}^7\mbox{Li}_2(v_i=0,j_i=0)$ at $E=0.928$ nK on
the scaling factor $\lambda$ of the nonadditive part of the
potential. Reproduced from Cvita\v s {\em et al.}
\cite{Cvitas:li3:2006}.}\ \label{sens-veq0}
\end{center}
\end{figure}

\begin{figure} [tbp]
\begin{center}
\includegraphics[width=65mm,angle=-90]{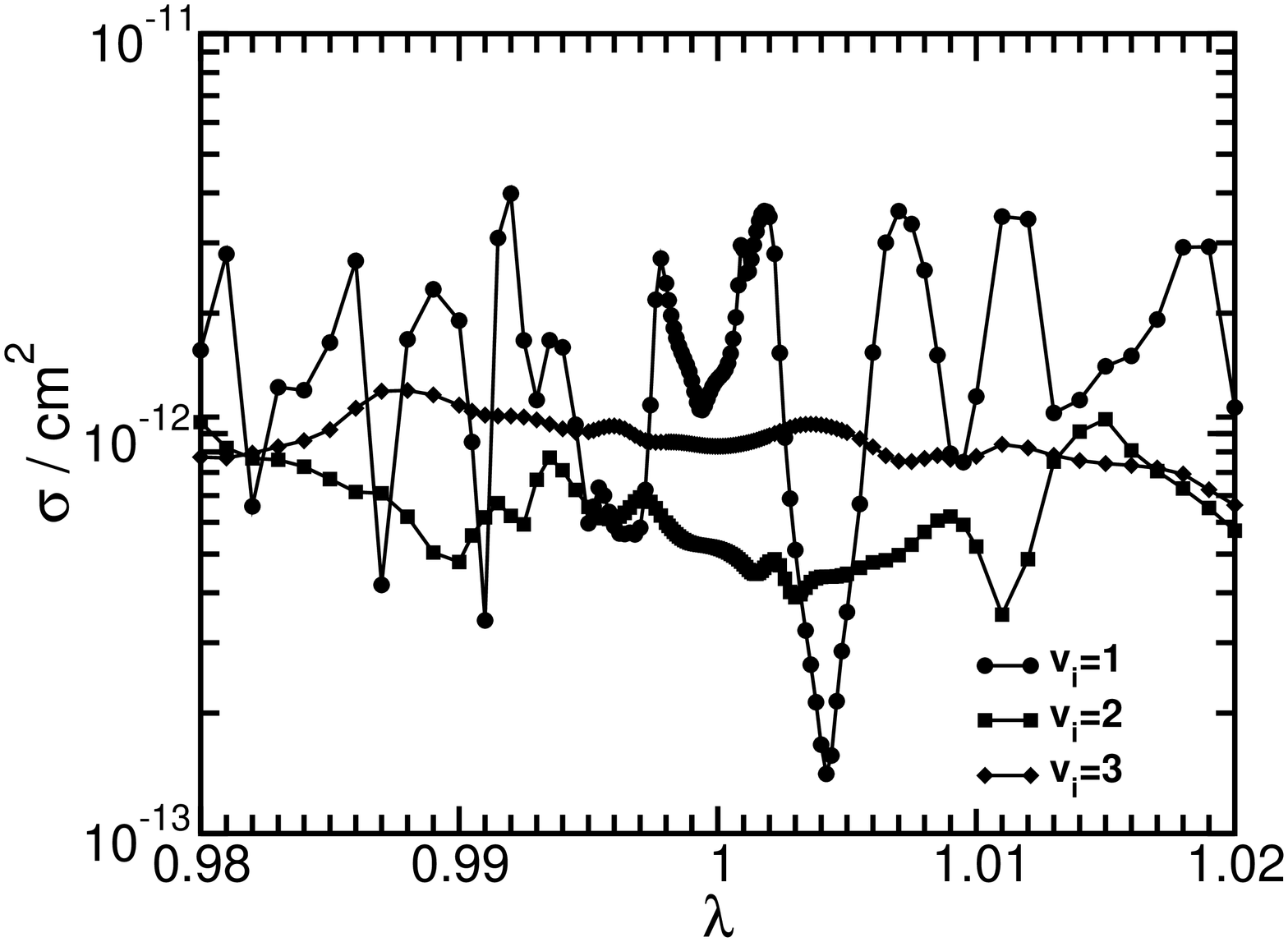}
\includegraphics[width=65mm,angle=-90]{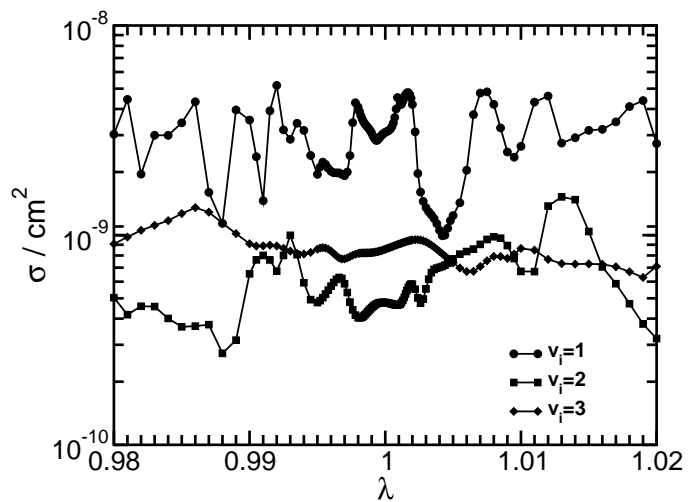}
\caption{Dependence of the total elastic (upper panel) and
inelastic (lower panel) cross sections for
$^7\mbox{Li}+\mbox{}^7\mbox{Li}_2(v_i,j_i=0)$ for $v_i=1$, 2 and 3
and $E=0.928$ nK on the scaling factor $\lambda$ of the
nonadditive part of the potential. Reproduced from Cvita\v s {\em
et al.} \cite{Cvitas:li3:2006}.}\ \label{sens-vgt0}
\end{center}
\end{figure}

The sensitivity of the cross sections to details of the potential
energy surface is of great importance. Sold\'an {\em et al.}\
\cite{Soldan:2002} showed that including the nonadditive part of
the interaction potential changed both elastic and inelastic cross
sections for Na + Na$_2\ (v=1)$ by more than a factor of 10.
Qu\'em\'ener {\em et al.}\ \cite{Quemener:2004} investigated this
in more detail: they introduced a scaling factor $\lambda$ to
multiply the nonadditive term, so that $\lambda=0$ corresponds to
a pairwise-additive potential and $\lambda=1$ to the full
nonadditive potential. They then investigated cross sections as a
function of $\lambda$ for initial $v=1$, 2 and 3. They found that
elastic and inelastic cross sections varied by a factor of 10 for
$v=1$ for variations of $\lambda$ as small as 0.01 either side of
$\lambda=1$. However, the variations became considerably smaller
for $v=2$ and 3. Cvita\v s {\em et al.}\ \cite{Cvitas:li3:2006}
have investigated similar effects for Li + Li$_2$ at rather higher
energy resolution. For $v=0$, the elastic cross sections show very
sharp structure as a function of $\lambda$ as shown in Fig.\
\ref{sens-veq0}, caused by poles in the scattering length each
time there is a bound state at zero energy. However, the structure
for $v>0$ is weaker for both elastic and inelastic cross sections
as shown in Fig.\ \ref{sens-vgt0}. This is due to a general effect
discussed by Hutson \cite{Hutson:res:2006}: in the presence of
inelastic scattering, the poles in scattering lengths at the
positions of zero-energy resonances are suppressed, and the
suppression increases with the degree of inelastic scattering.

\subsection{Differential cross sections}

\begin{figure} [tb]
\begin{center}
\includegraphics[width=85mm]{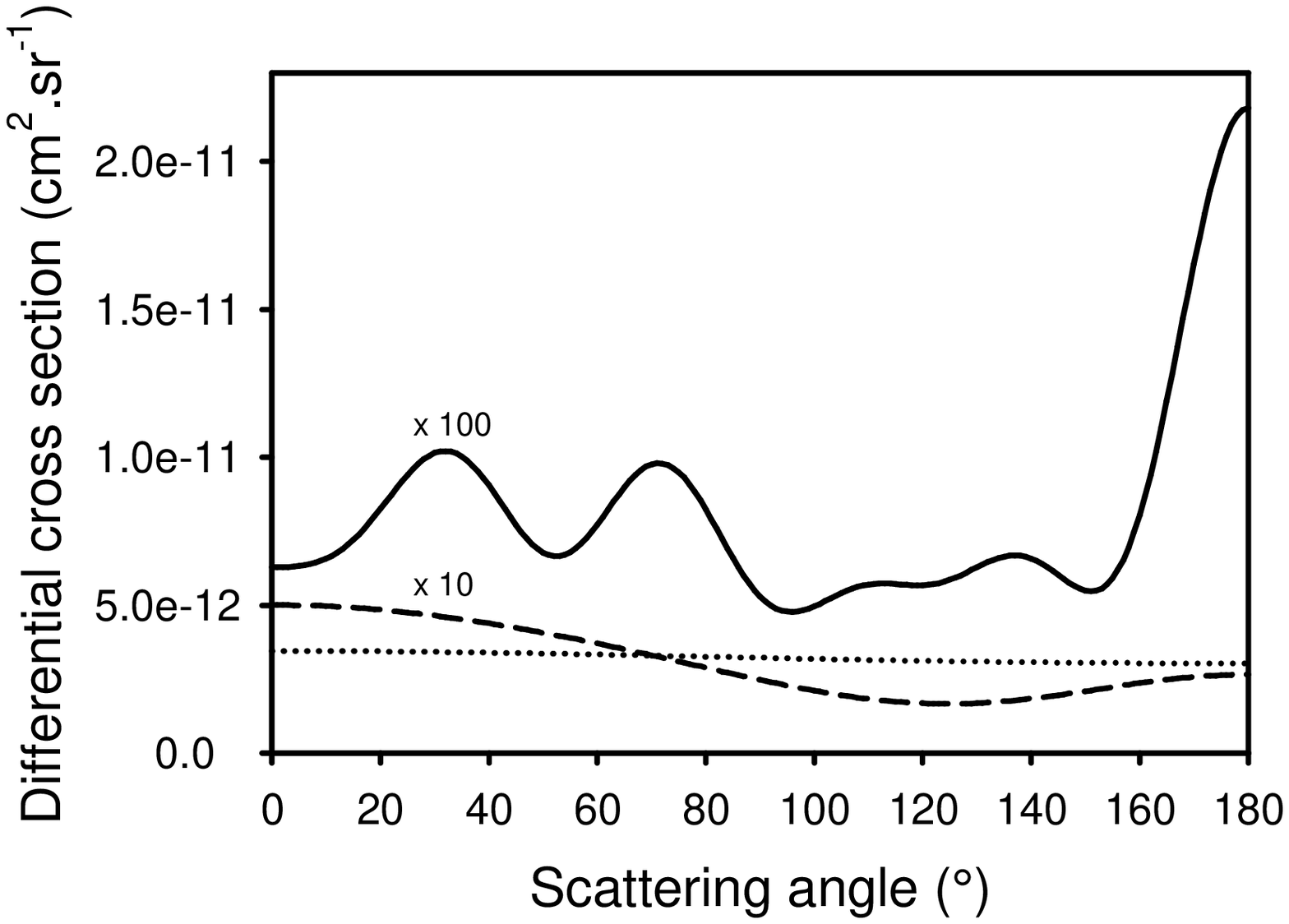}
\caption{Differential cross section for inelastic scattering of K
+ K$_2\ (v=1)$ at 1~$\mu$K (dotted line), 100~$\mu$K (dashed line)
and 0.1~mK (solid line). Reproduced from Qu\'em\'ener {\em et
al.}\ \cite{Quemener:2005}} \label{dcs}
\end{center}
\end{figure}

At very low energies, cross sections are completely dominated by
the $l=0$ partial wave. Under these circumstances, the scattering
is completely isotropic and the differential cross sections are
featureless. However, as the energy increases and higher partial
waves start to contribute, angular structure appears. Low-energy
scattering thus offers the opportunity to study the onset of
angular behaviour in reactive cross sections. The way that the
angular behaviour develops is shown for K + K$_2\ (v=1)$ in Fig.\
\ref{dcs}. At 1 $\mu$K the scattering is completely isotropic, but
for 100 $\mu$K some angular structure arising from interference
between $l=0$ and 1 is evident. At 0.1 mK partial waves up to
$l=5$ contribute and several peaks emerge.

\subsection{Heteronuclear molecules}

Heteronuclear molecules are particularly interesting because they
offer the possibility of studying {\it reactive} collisions
separately from inelastic collisions. Homonuclear molecules that
are formed in their lowest vibration-rotation state are stable to
collisions. For heteronuclear molecules, however, the situation is
more complicated. Even molecules in their ground rovibrational
states may not be stable against collisions. For example, the
spin-polarized reaction
\begin{equation}
^6\textrm{Li}^7\textrm{Li}(v=0,j=0) + {}^7\textrm{Li} \rightarrow
{}^6\textrm{Li} + {}^7\textrm{Li}_{2}(v=0,j=0) \label{react1}
\end{equation}
is exothermic by 1.822~K because of the difference between the
zero-point energies of the two dimers. However, the process
\begin{equation}
^6\textrm{Li}^7\textrm{Li}(v=0,j=0) + {}^6\textrm{Li} \rightarrow
{}^7\textrm{Li} + {}^6\textrm{Li}_{2}(v=0,j=1) \label{react2}
\end{equation}
cannot occur at collision energies below 2.643\,K because of the
combined effects of zero-point energy and the need to form
$^6$Li$_2$ in $j=1$ or higher to satisfy fermion symmetry
requirements.

\begin{figure} [tb]
\begin{center}
\includegraphics[width=70mm,angle=-90]{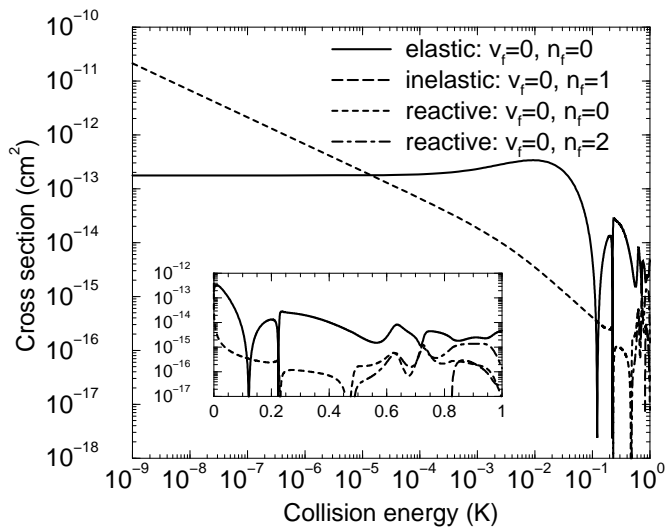}
\caption{Elastic and reactive s-wave cross sections for $^7$Li +
$^6$Li$^7$Li$(v=0,j=0)$. The inset shows the higher-energy cross
sections on a log-linear scale, with axes in the same units as the
main plot. Reproduced from Cvita\v s {\em et al.}\
\cite{Cvitas:hetero:2005}.} \label{xsec-7-67}
\end{center}
\end{figure}

Cvita\v s {\em et al.}\ \cite{Cvitas:hetero:2005} have
investigated the process (\ref{react1}) and the resulting elastic
and reactive cross-sections are shown in Fig.\ \ref{xsec-7-67}. It
may be seen that the reactive scattering dominates over elastic
scattering below 10 $\mu K$. The low-temperature reactive rate
coefficient is only $4.7 \times 10^{-12}$ cm$^3$ s$^{-1}$, which
is reduced by a factor of about 50 from those typical for
vibrational relaxation in the homonuclear case. Cvita\v s {\em et
al.}\ attributed the difference to the fact that there is only a
single open channel for reaction (\ref{react1}).

The results obtained in Ref.\ \onlinecite{Cvitas:hetero:2005} have
important implications. There is interest in producing a quantum
gas of $^6$Li$^7$Li in its ground rovibronic state in an ultracold
mixture of $^6$Li and $^7$Li atoms. In order to stabilize the
molecular cloud against two-body trap losses induced by the
reactive process (\ref{react1}), the remaining atomic $^7$Li would
have to be removed quickly after ground-state molecule production,
so that just the two-species fermionic mixture of
$^6$Li$^7$Li$(v=0,j=0)$ molecules and $^6$Li atoms is left in the
trap. The $^6$Li cloud could be removed as well, but it might be
advantageous to keep it in the trap. Elastic s-wave collisions
between fermionic $^6$Li$^7$Li molecules will be strongly
suppressed, but low-energy collisions with $^6$Li can result only
in elastic scattering and might be used to achieve sympathetic
cooling of the molecules.

\subsection{Further extensions}

Our work on the quantum dynamics of collisions of alkali metal
dimers has so far been restricted in several ways. We have
focussed on collisions of molecules in low vibrational states, for
systems involving three chemically equivalent atoms. We have
restricted ourselves to collisions of spin-stretched atoms and
molecules, for which doublet electronic states do not contribute.
We have neglected hyperfine structure, and worked in zero applied
field.

Extending the calculations to handle heteronuclear systems is
relatively straightforward, though a considerable amount of work
is needed to develop potential energy surfaces for each system of
interest. Dynamical calculations are more expensive for heavier
atoms and for systems of lower symmetry (and the calculations
described here already push the limits of current computers).
Extending the calculations to higher vibrational states is also
mostly a matter of computer time, though true long-range states
very near dissociation may be difficult or impossible to converge
with our current scattering methods.

Including the effects of nuclear spin and magnetic fields is a
very difficult task, though an important one if we are to explore
atom-molecule Feshbach resonances and use them to control
molecular interactions in the same way as atomic interactions.
Collisions of atoms and molecules in non-spin-stretched states
will be particularly challenging, because they will involve
doublet surfaces as well as quartet surfaces, and for alkali metal
trimers the doublet surfaces exhibit conical intersections and
geometric phase effects \cite{Kendrick:geometric:2003,
Althorpe:2006} that considerably complicate the dynamics.

\section{Conclusions}

This article focussed on theoretical studies of collisions between
spin-polarized alkali metal dimers and atoms, which are crucial in
experiments designed to form ultracold molecules in low-lying
vibrational states. Colliding dimers can undergo very fast
barrierless chemical reactions. As a result, vibrationally excited
molecules undergo very fast vibrational relaxation, with rates
usually in excess of $k_{\rm inel} = 10^{-10}$ cm$^3$ s$^{-1}$. At
temperatures above about 1 mK, where several partial waves
contribute, the rates are approximately given by a statistical
Langevin capture model. At lower temperatures, however, the
reactions enter a regime governed by Wigner threshold laws and a
full quantum-dynamical treatment is essential to calculate the
rates. In this regime the results are very sensitive to details of
the triatomic potential energy surfaces, though the sensitivity
decreases for excited vibrational levels. Isotopically
heteronuclear molecules can often undergo exothermic reactions
even from their ground vibrational states, because of the
difference in zero-point energy between reactants and products.

Prospects for the future include the production of
quantum-degenerate gases of ground-state molecules, which will be
stable to collisions and offer a wealth of new possibilities for
quantum control. Heteronuclear molecules are particularly
interesting, because they can have substantial dipole moments in
long-range states. Dipolar quantum gases offer a new range of
novel properties, and ultracold polar molecules also have
potential applications in quantum computing and in studying
fundamental physical properties such as parity violation and the
electron dipole moment.

\section{Acknowledgments}

PS acknowledges support from the Ministry of Education of the
Czech Republic (grant no. LC06002).

\bibliography{../../all}

\end{document}